\documentclass[letterpaper,twocolumn,10pt]{article}
\usepackage{usenix,epsfig,endnotes}
\usepackage{graphicx}
\usepackage{balance}  
\usepackage{color}
\usepackage{pgfplots}
\usepackage{pgfplotstable}
\usepackage[]{subcaption}
\usepackage{url}
\usepackage[font=small]{caption}
\DeclareCaptionType{copyrightbox}
\usepackage{tikz}
\usepackage{environ}
\usetikzlibrary{fit}
\usetikzlibrary{calc}
\usetikzlibrary{patterns}
\usetikzlibrary{shapes.multipart, arrows}
\usetikzlibrary{decorations.pathreplacing}
\usepackage{pgfplots}
\usepackage{amssymb}
\usepackage{amsmath}
\usepackage{multirow}
\usepackage{relsize}
\usepackage[titletoc]{appendix}
\usepackage{xcolor,colortbl}
\usepackage[font={scriptsize}]{subcaption}
\usepackage{authblk}
\usepackage[linesnumbered,ruled,vlined]{algorithm2e}
\SetAlgorithmName{Algorithm}{Algorithm}{Algorithm}
\IncMargin{0.5em}
\SetCommentSty{textnormal}
\SetNlSty{}{}{:}
\SetAlgoNlRelativeSize{0}

\newcommand{\ustore}{{UStore}}
\newcommand{\uobject}{{\tt UObject}}
\newcommand{\acuobject}{{\tt ACUObject}}
\newcommand{\unode}{{\tt UNode}}
\begin{document}\sloppy

\title{UStore: A Distributed Storage With Rich Semantics}
\author[1]{Anh Dinh}
\author[1]{Ji Wang}
\author[1]{Sheng Wang}
\author[2]{Gang Chen}
\author[1]{Wei-Ngan Chin}
\author[1]{Qian Lin}
\author[1]{Beng Chin Ooi}
\author[1]{Pingcheng Ruan}
\author[1]{Kian-Lee Tan} 
\author[1]{Zhongle Xie}
\author[1]{Hao Zhang}
\author[3]{Meihui Zhang}
\affil[1]{National University of Singapore}
\affil[2]{Zhejiang University}
\affil[3]{Singapore University of Technology and Design}
\maketitle

\begin{abstract}
Today's storage systems expose abstractions which are either too low-level (e.g., key-value store, raw-block
store) that they require developers to re-invent the wheels, or too high-level (e.g., relational databases,
Git) that they lack generality to support many classes of applications. In this work, we propose and
implement a general distributed data storage system, called \ustore, which has rich semantics. \ustore\ delivers
three key properties, namely immutability, sharing and security, which unify and add values to many classes of
today's applications, and which also open the door for new applications. By keeping the core properties within
the storage, \ustore\ helps reduce application development efforts while offering high performance at hand. The
storage embraces current hardware trends as key enablers. It is built around a data-structure similar to that
of Git, a popular source code versioning system, but it also synthesizes many designs from distributed systems
and databases. Our current implementation of \ustore\ has better performance than general in-memory key-value
storage systems, especially for version scan operations. We port and evaluate four applications on top
of \ustore: a Git-like application, a collaborative data science application, a transaction management application, and
a blockchain application. We demonstrate that \ustore\ enables faster development and the \ustore-backed applications
can have better performance than the existing implementations.  \end{abstract}

\vspace{1em}

\section{Introduction}
Application developers today can choose from a vast array of distributed storage systems. As storage costs are going
down drastically, storage systems differentiate themselves by the levels of abstractions offered to the applications. At
one extreme, key-value stores such as Dynamo~\cite{dynamodb}, Voldemort~\cite{voldermot}, Redis~\cite{redis},
Hyperdex~\cite{hyperdex} provide simple interface to build highly available and scalable applications. However, many
systems have more complex data models than simple key-value records, for instance social graphs or images, such that to
implement them on top of a key-value storage requires re-building complex software stacks and therefore risks
re-inventing the wheel. At the other extreme, relational databases are highly optimized, but enforce strict relational
data models which limit the range of applications, and the ACID guarantees limit their scalability. Currently, we see a
shift towards more structured storage systems, for examples, TAO~\cite{tao}, MongoDB~\cite{mongodb},
PNUTS~\cite{pnuts}, LogBase~\cite{logbase}, which offer more scalability, but their data
models are specific to the given domain, therefore they do not generalize well to other domains. Amid these
choices, we ask the question whether there is still a gap to be filled by another storage system, and if
there is, what would be the right level of abstraction? 

We observe that many recent distributed applications share three core properties. First, {\em data
immutability}, in which data never changes and any update results in a new version, is being exploited to build popular
version control systems like Git~\cite{git}. Immutability also plays a key role in many data intensive
systems~\cite{immutability} wherein it helps simplify the design for fault tolerance and replication.
MapReduce~\cite{mapreduce} and Dyrad~\cite{dryad}, for instance, split computation tasks into smaller units each taking
immutable input and storing immutable output to HDFS or GFS. Immutability in these settings make it easy to handle
failures by simply restarting the units without complex coordination. Second, {\em data sharing} is the key property in
the growing number of social network and collaborative applications, driven by the growth of user-generated data and the
need for collaborative analytics~\cite{adam_sigmod15}. Early peer-to-peer content sharing systems~\cite{oceanstore,
bittorrent} are being replaced by more centralized alternatives~\cite{datahub} with different data
models and collaborative workflows. For example, DataHub~\cite{datahub} exposes a dataset model with infrequent updates,
whereas GoogleDocs assumes text documents and real-time updates.  Third, {\em data security} is
becoming increasingly important due to data breaches arising from insider threats (NSA, Target, Sony hack, for example),
and due to the vulnerability of third-party cloud providers to attacks.
For confidentiality protection, systems such
as CryptDb~\cite{cryptdb} and M2R~\cite{m2r} enable database and big-data analytics tasks to be performed
over encrypted data,
by employing novel encryption schemes and trusted hardwares. Blockchain systems like Bitcoin~\cite{bitcoin},
Ethereum~\cite{ethereum} and Hyperledger~\cite{hyperledger} ensure integrity of the data stored on a shared ledger
even under Byzantine failures. 

Given these trends, we argue that there are clear benefits in unifying data immutability, sharing and security in a
single storage system. By focusing on optimizing these properties in the storage, one immediate value is to reduce
development efforts for new applications that need any combination of these properties, as they are provided out of the
box. In addition, existing applications can be ported to the new storage, thereby achieving all three properties at the
same time and with better performance.  

In this paper, we propose a new distributed storage system, called \ustore, that accomplishes the following goals.
First, it has rich semantics: the storage supports data immutability, sharing and security. Second, it is flexible:
performance trade-offs can be tuned, and it can be easily extended to accommodate new hardware or application's needs.
Third, it is efficient and scalable: it offers high throughput and low latency, and it can scale out
to many servers.  Existing systems fall shorts at accomplishing all three goals at the same time. For instance, temporal
databases~\cite{6544906} or HDFS provide only immutability, whereas P2P systems~\cite{bittorrent} focus mainly on sharing.
Git~\cite{git} and Datahub~\cite{datahub} implement both immutability and sharing, but forgo security.
Furthermore, Git is designed for P2P environment, thus it is limited in efficiency~\cite{git-scale}.  Datahub is
restricted to relational data models (hence, it is inflexible) and has not been shown to scale to multiple servers.
Several systems like SPORC~\cite{sporc} and Bitcoin~\cite{bitcoin} support all three properties, but they are not
general and efficient enough for other applications beside collaborative text editing and crypto-currency.   

\ustore's data model is based on a novel data structure called \uobject, which is identified by a unique key and its
content is a direct acyclic graph of \unode\ objects. Each \unode\ in the graph is identified by a unique version, its
value contains the data, and connections between them represent derivation relation between different versions of the
data. An update to \uobject\ creates a new version in the graph, and the complete version history can be traced by following
the backward pointers. This is similar to the commit history in Git, but unlike Git, \ustore\ partitions and replicates
\unode\ objects over multiple nodes for better read and write performance. \ustore's partitioning scheme is locality-aware,
in the sense that related versions are likely to be grouped together in the same node. Together with caching, and native
support for Remote Direct Memory Access (RMDA), \ustore\ can deliver high performance on operations that scan historical
versions. \ustore\ supports non-realtime collaborative workflows by providing a publish/subscribe channel with which
users can be notified when there are new versions of the objects of interest. It guarantees integrity of both the data
and version history against untrusted providers by using tamper-evident version numbers which are similar to hash
pointers in blockchains. Furthermore, it allows for flexible and fine-grained sharing policies based on the entire
\uobject\ or a set of \unode\ objects. \ustore\ provides a number of parameters for tuning performance trade-offs
between the access operations, storage capacity and overall availability guarantees. Finally, \ustore\ supports
push-down semantics by allowing user-defined logics for data compression, and for detecting and merging of data
conflicts.      

We implement four applications on top of the current in-memory implementation of \ustore. The first application is an
extension of Git for distributed settings, in which multiple users collaborate on a single repository. Another
application implements the relational data model and data science workflows as supported in Datahub~\cite{datahub}. The
third application is a transaction management protocol based on TARDiS~\cite{tardis} that supports weak consistency with
branch-and-merge semantics. Finally, we implement a simple private blockchain application based on
Ethereum~\cite{ethereum} which supports analytical queries on the blockchain data. Each application is implemented in
fewer than $1300$ lines of code, demonstrating that \ustore\ enables fast application development. We then benchmark
\ustore's performance individually, and evaluate the four applications against systems that support the same operations.
The results show that \ustore's performance is comparable to Redis in basic read/write operations, and is better than
Redis in scanning operations. The Git-like application achieves up to $3$ orders of magnitude lower latency than Git in
versioning operations (commit and checkout). The collaborative data science application achieves up to $10$x lower
latency than Decibel~\cite{decibel} in 3 out of 4 popular queries. The transaction management application reduces the
number of states accessed by $2$ orders of magnitudes compared to TARDiS. The blockchain application outperforms
Ethereum by $2$ orders of magnitudes in 4 out of 5 queries. 




In summary, in this paper we make the following contributions:
\begin{itemize}
\item We identify common trends in today's distributed applications and argue for the need of a distributed storage
targeting a large classes of existing and emerging applications. 
\item We design and implement \ustore, a flexible, efficient and scalable distributed storage with three core
properties: data immutability, sharing and security.
\item We benchmark \ustore\ against Redis, showing comparable performance for basic operations and significant
improvement for scan operations.  We  implement four applications on top of \ustore, namely Git, collaborative data
science, transaction management and blockchain. We evaluate them against systems supporting the same operations, and
demonstrate that \ustore\ improve the applications' performance while reducing development efforts. 
\end{itemize}

In the next section, we motivate \ustore\ by discussing the trends in distributed applications and the
challenges faced by these systems. We present the detailed design in Section~\ref{sec:design}, and describe our
implementation of four applications in Section~\ref{sec:newapp}. We report the performance of \ustore\ and of its
applications in Section~\ref{sec:evaluation}. We discuss \ustore's current states and future work in
Section~\ref{sec:discussion}, before concluding in Section~\ref{sec:conclusion}. 

\section{Related Work and Motivations}
\label{sec:motivation}
In this section, we discuss several trends in distributed systems that underpin many interesting applications, including
version control, data versioning, collaboration and security-aware applications, and related work. Table~\ref{tab:list} lists research and
open source systems along three common properties: immutability, sharing and security. We then review new hardware
capabilities that are key enablers for next-generation distributed systems.  


\begin{table}
\resizebox{\columnwidth}{!}{
  \begin{tabular}{|l|l|l|l|}
  \hline
  {\bf System} & {\bf Immutability} & {\bf Sharing} & {\bf Security} \\ \hline
  GFS/HDFS~\cite{gfs} & $\checkmark$ & & \\ \hline
  RDD~\cite{Zaharia:2012:RDD:2228298.2228301} & $\checkmark$ & & \\ \hline
  Datomic~\cite{datomic} & $\checkmark$ & & \\ \hline
  LogBase~\cite{logbase} & $\checkmark$ & & \\ \hline
  Bittorent~\cite{bittorrent} & & $\checkmark$ & \\ \hline
  Dropbox~\cite{dropbox} & & $\checkmark$ & \\ \hline
  Tahoe LAFS~\cite{lafs} & & $\checkmark$ & $\checkmark$ \\ \hline
  Datahub~\cite{datahub} & $\checkmark$ & $\checkmark$ & \\ \hline
  Git~\cite{git} & $\checkmark$ & $\checkmark$ & \\ \hline 
  Irmin~\cite{irmin} & $\checkmark$ & $\checkmark$ & \\ \hline 
  Noms~\cite{noms} & $\checkmark$ & $\checkmark$ & \\ \hline
  Pachyderm~\cite{pachyderm} & $\checkmark$ & $\checkmark$ & \\ \hline
  Ori~\cite{orifs} & $\checkmark$ & $\checkmark$ & \\ \hline 
  SUNDR~\cite{sundr} & $\checkmark$ & $\checkmark$ & $\checkmark$ \\ \hline 
  Bitcoin~\cite{bitcoin} & $\checkmark$ & $\checkmark$ & $\checkmark$ \\ \hline 
  Ethereum~\cite{ethereum} & $\checkmark$ & $\checkmark$ & $\checkmark$ \\ \hline 
  \hline
  {\bf \ustore} & $\checkmark$ & $\checkmark$ & $\checkmark$ \\ \hline
  \end{tabular}
}
\caption{Systems built around data immutability, sharing and security.}
\label{tab:list}
\end{table}
\subsection{Immutability}
Git~\cite{git} is a widely used open-source distributed version control system (DVCS), which outperforms other VCS (such
as Subversion, CVS, Perforce) due to its unique features like cheap local branching, convenient staging areas, and
multiple workflows. Fundamental to Git's design is data immutability, that is, all changes committed to Git are
permanently archived in version histories. Viewed as an append-only key-value store, Git allows efficient tracking of the
entire version history. Furthermore, it is easy in Git to compare, merge and resolve conflicts over branches. Git can
automatically resolve many conflicts arising from source code versioning, and only notifies users for conflicts it cannot
resolve. Git enables {\em offline} collaboration models in decentralized, P2P settings in which each user has a complete
copy of the repository. 

Beside Git, we observe immutability in many other data-oriented applications. In particular, massively parallel
systems such as MapReduce and Dryad are based on immutable inputs stored in HDFS or GFS, which greatly simplifies failure
handling by restarting the failed tasks. Similarly, Spark~\cite{Zaharia:2012:RDD:2228298.2228301} is based on Resilient Distributed Datasets (RDDs) abstraction,
which are indeed immutable datasets tracking the operation history. Other examples of immutability include  
LSM-like data storages such as HBase~\cite{hbase}, in which immutability enables superior write
throughput for ingesting updates while simplifying failure recovery.

One particular manifestation of immutability in data management systems is {\em data versioning}, which has been
employed for tolerating failures, errors and intrusions, and for analysis of data modification history.
ElephantFS~\cite{hotos99:Santry} is one of the first file systems with built-in multi-version support.
Later systems like S4~\cite{osdi00:Strunk}, CVFS~\cite{fast03:Soules}, RepareStore~\cite{dsn03:Zhu}
and OceanStore~\cite{oceanstore}, improve the early design by maintaining all versions in full scope and upon
each update operation.  Most of these systems use journal/log-structure file system (e.g.,
SpriteLFS~\cite{tocs92:Rosenblum}) as the underlying storage, because they leverage the latter's high performance in
append-only workloads.  In databases, data versioning techniques are used for transactional data access.
Postgres~\cite{vldb87:Stonebraker}, for example, achieved
performance comparable to other database systems without versioning support.  Fastrek~\cite{icde05:Chiueh} enhanced
Postgres with intrusion tolerance by maintaining an inter-transaction dependency graph based on the versioned data, and
relying on the graph to resolve data access conflicts.

\subsection{Sharing and Collaboration}
The exponential growth of data can be attributed to the growth in user-generated, machine-generated data and the need to
log and collect data for future analytics. Before the data can be meaningfully used, the owners must be able to share
their data among each other and among different systems. Past~\cite{past} and Bittorrent~\cite{bittorrent}, for
examples, are optimized for object discovery, durability, availability and network bandwidth. Data sharing
among users is fundamental to recent social network platforms, for which many techniques have been
developed to optimize both throughput and latency~\cite{haystack}. 

Efficient data sharing makes it possible to implement various collaboration models. Unlike classic multi-user,
time-sharing systems like databases which give a user the illusion of owning the entire system, collaborative systems
explicitly provide different views to different users, and synchronization between different views are directly observed
and controlled by the users.  Most collaborative systems expose file system interface where a set of files can be mounted
locally from a remote server.  Dropbox, NFS and CIFS, for example, assume centralized servers, whereas
Git~\cite{git}, IPFS~\cite{ipfs}, Ori~\cite{orifs} work in decentralized settings. These systems
employ light-weight synchronization techniques such as versioning and content-addressable files in order to minimize
synchronization cost. They can be characterized by their supported workflows: from infrequent updates (version control
systems like Git), frequent updates (shared file systems like Dropbox), to real-time updates (document editing systems
like  GoogleDocs). Recent systems such as DataHub~\cite{datahub} exposes a {\em dataset} interface to
support collaborative big-data workloads. Datahub targets scientific domains wherein multiple users and teams perform
data-intensive computations on shared data~\cite{adam_sigmod15}, for which existing databases or version control systems
are inadequate.     


\subsection{Security}

There is an inherent threat from moving data into the hand of untrusted
parties, i.e. cloud providers. Recent high-profile data breaches and system attacks (NSA, Target,
Sony hack, for examples) further demonstrate the challenges in protecting data from insider threats. Protecting data
confidentiality can be readily implemented on existing cloud storage systems, by simply encrypting the data. However,
there is a need to perform computation on the encrypted data, for which systems like CryptDb~\cite{cryptdb}
employ homomorphic encryption schemes.
In order to support a rich set of database operations, these systems make strong assumptions on the data 
and security model, which may not hold in practice~\cite{edb}. Recent systems, namely Haven\cite{haven} and M2R~\cite{m2r},
rely on trusted hardware to deliver high security guarantee for general computations using
small trusted code base.  

In a collaborative setting with an untrusted provider, integrity protection refers to the ability to detect {\em forks}.
Specifically, the provider can present different sequences of updates to the shared state to different users,
thereby forking multiple views. SUNDR~\cite{sundr} is the first system to provide fork-consistent file systems, meaning
that if the server presents two users with different views, these users can either never see each other's view, or they
can detect that there is a fork.  Later works, such as Venus~\cite{venus} and Depot~\cite{depot} extended SUNDR to
improve performance and conflict resolutions.  In the decentralized setting where peers distrust each other, recent
blockchain systems achieve integrity for a global data structure resembling a ledger~\cite{bitcoin,ethereum,hyperledger}.
In these systems, users reach agreement via distributed consensus protocols which can tolerate certain numbers of
malicious adversaries. In public blockchain systems, in which peers can freely join and leave, the consensus protocol is
based on proof-of-work which gives each user a probability of updating the blockchain proportional to his computing
power. Recent proposals of private blockchains can achieve better performance by using cheaper consensus protocols such
as PBFT~\cite{pbft}. 

\subsection{Hardware Trend}
The increased availability of large memory has given rise to in-memory computing~\cite{Tan:2015:IDC:2814710.2814717, zhang2015memory}, from general computing frameworks like
Spark~\cite{Zaharia:2012:RDD:2228298.2228301}, to databases like
HyPer~\cite{kemper2011hyper} and SAP
HANA~\cite{6544906}, or data storage systems like RAMCloud~\cite{179821} and Redis~\cite{redis_cite}. In-memory systems
deliver low latency and thus can be used for real-time analytics. 

Beside memory, new hardware primitives such as Non-uniform Memory Access (NUMA)~{\cite{Maas:2013:BNI:2463676.2465342}},
Hardware Transactional Memory
(HTM)~{\cite{leis2014exploiting}}, Remote Direct Memory Access (RDMA) networking~\cite{kalia2014rdma}, Non-Volatile
memory (NVM)~\cite{DeBrabant2014vldb}, etc.  offer new opportunities to improve system performance by leveraging the
hardware. However, changes in hardware often require re-examining the existing designs in order to fully exploit the hardware
benefits. For example, HyPer~\cite{Leis_NUMA} proposes a new query evaluation framework to overcome overheads with
non-NUMA-aware data access. Pilaf~\cite{Mitchell:2013:UOR:2535461.2535475}, HERD~\cite{kalia2014rdma}, and
FaRM~\cite{dragojevic2014farm} propose enhancement to existing key-value storage and transactional systems to fully
exploit RDMAs. New trusted computing capabilities such as the new Intel SGX are being employed by privacy-preserving
systems such as Haven~\cite{haven}, M2R~\cite{m2r} to significantly reduce overhead of computing on encrypted data. 


\section{UStore Design}
\label{sec:design}
\begin{figure}
\includegraphics[width=0.48\textwidth]{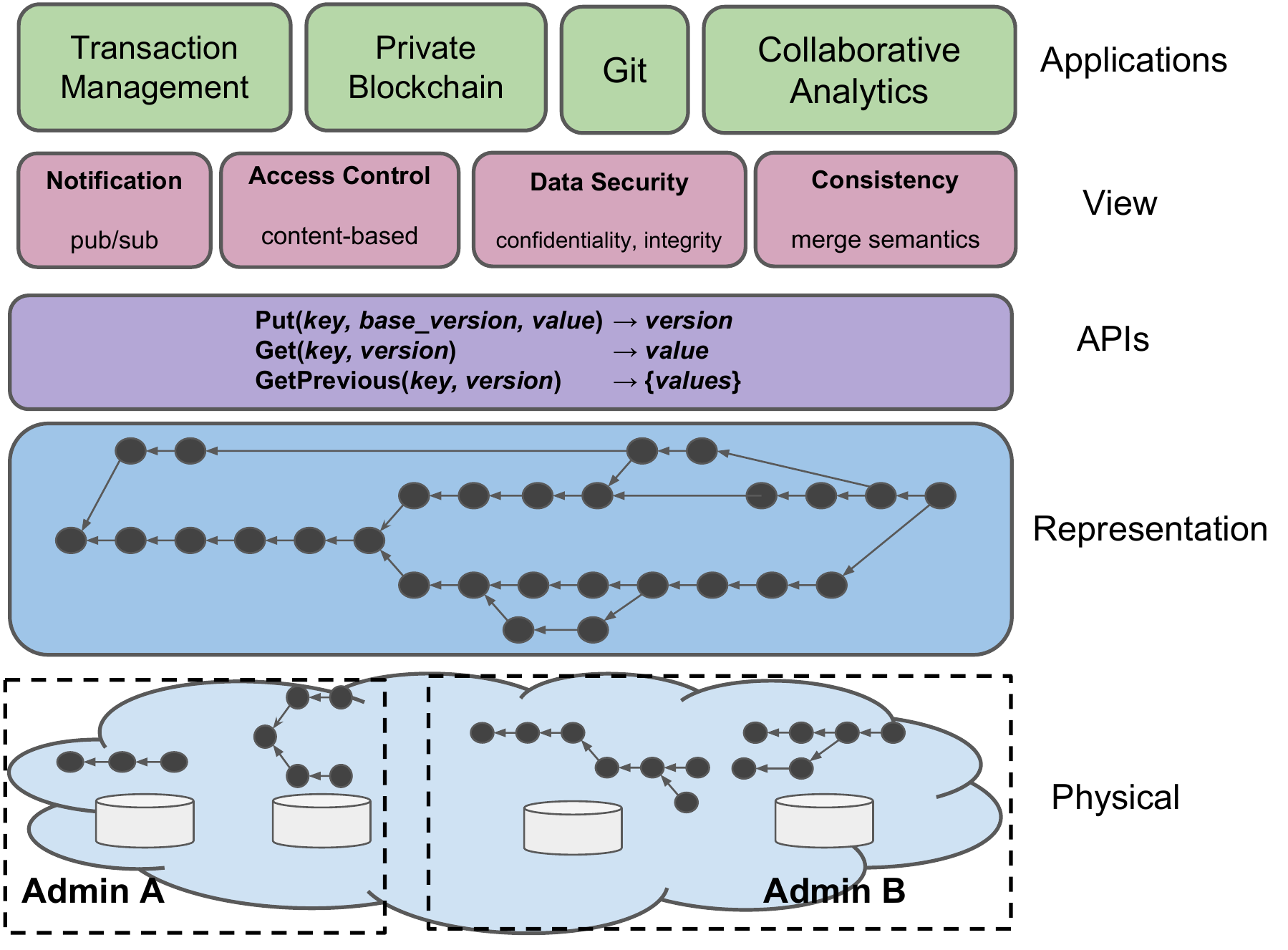}
\caption{\ustore\ stack consists of multiple layers. The data representation layer constitutes the {\em narrow waist}.}
\label{fig:stack}
\end{figure}
In this section, we present the detailed design of \ustore. The system follows a layered design, with the
narrow waist being the core abstraction called \uobject. We first discuss the high level design goals, and then describe
how we achieve them at different layers of the system.  

\subsection{Overview}
Inspired by the software and hardware trends discussed in the previous section, we design \ustore\ with three
high-level goals.

{\bf G1: Rich semantics.} \ustore\ provides data immutability, data sharing, and data security out of the box.  Once
data is inserted, it remains unchanged. The data can be shared among applications and users in fine-grained manner.
Finally, integrity of the data and of the update history are protected against untrusted cloud providers.  

{\bf G2: Flexibility.} \ustore\ APIs give its applications freedom to configure and combine the immutability, sharing and security properties.
It also allows user to push down application semantics via a number of well-defined interfaces. 

{\bf G3: Efficiency \& scalability.} \ustore\ delivers high throughput, low latency for its data access operations,
and it can scale out to many nodes. 




Figure~\ref{fig:stack} shows multiple layers of \ustore's stack. At the bottom, the physical layer is responsible
for storing, distributing and replicating data over many servers. We consider settings in which servers belong
to a number of independent administrative domains, i.e. they are mutually distrustful. The next layer
contains \ustore's core data structure which bears some similarity to Git's.
In particular, it implements immutability with support for branching and merging. \ustore\ 
differentiates from Git in the structure and operational semantics of each data value. The distinction to Git
becomes more apparent at the next two layers. The APIs layer exposes access operations which are specific to a
version of data. The view layer adds fine-grained access control, security, customizable consistency model and
a notification service.   Finally, applications such as transaction management, blockchains, data science
collaboration, etc. can exploit \ustore's rich semantics by building directly on the APIs and the views from
the lower layers. 

The Git-like data structure embedded at the second layer makes up the narrow waist of the design. In other
words, \ustore\ enforces a single representation and unchanged semantics of this data structure, but allows
for different implementations of the physical, view and application layers. By fixing the representation, \ustore\ can accommodate
  innovations at the physical layer and changes in application requirements. This layered design achieves the
  three goals as follows.  First, G1 is realized by the data representation and view layer. Second,
  flexibility (G2) is achieved by exposing parameters and application hooks at the physical layers for
  specifying constraints on resources, on distribution and replication strategies. In addition, \ustore\ lets
  the applications overwrite the consistency view to implement their own models. Third, \ustore's high
  performance (G3) comes from the careful use of the available hardware at the physical layer. 
\subsection{Abstraction and APIs}

\ustore\ is based on a novel abstraction, called \uobject, for reading and writing data.  Each \uobject\ 
manages all data related to a specific \emph{key} and supports retrieval of existing values and tracking of
version history.

\subsubsection{\uobject}
\label{subsub:uobject}
\begin{figure}
\centering
{\includegraphics[width=0.4\textwidth]{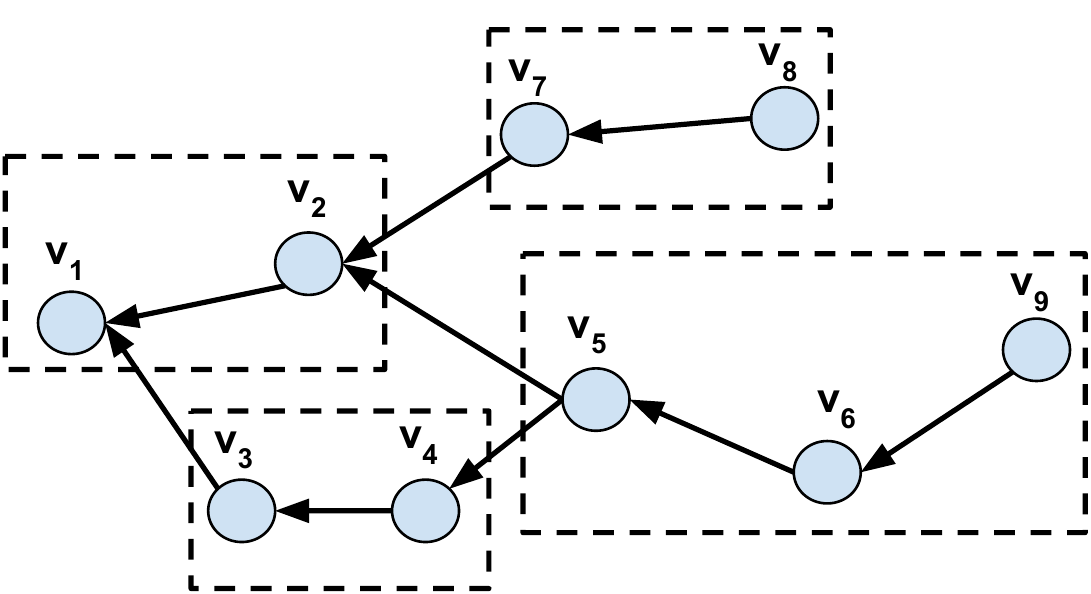}}
\caption{\uobject\ data structure.}
\label{fig:uobject}
\end{figure}

\begin{figure}
\centering
{\includegraphics[width=0.4\textwidth]{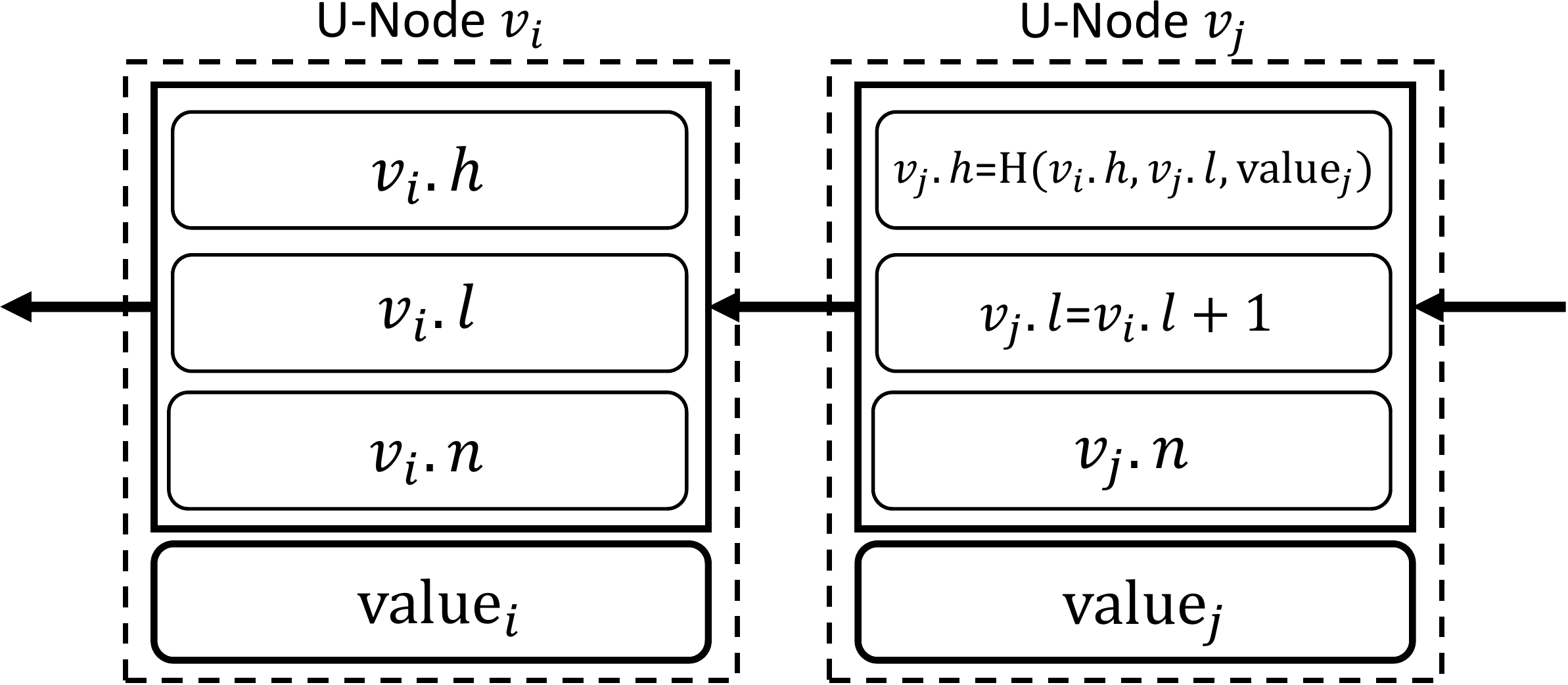}}
\caption{\unode\ data structure.}
\label{fig:unode}
\end{figure}

A \uobject\ is identified by a unique key, and comprises a directed acyclic graph (DAG), as shown in
Figure~\ref{fig:uobject}.  Each node in this graph, called {\unode}, contains a pair of data: a unique
\emph{version number} and the corresponding \emph{value} of that key, as shown in Figure~\ref{fig:unode}.  The
connections between {\unode}s represents derivation relations among different versions of the data.  In other words, a
\uobject\ contains all historical information of a key. Its concrete data structure is as follows: 

\begin{verbatim}
struct UObject {
    string key
    collection<UNode> versions 
}

struct UNode {
   version_t version;
   string value;
   version_t prev_1, prev_2; 
}

struct version_t {
  string h; 
  int l; 
  string n; 
}
\end{verbatim}

We have no restrictions on how each version can be extended.  That is, a version can be extended multiple
times, leading to independent branches, for example $v_2$ is extended to $v_5, v_7$ in Figure~\ref{fig:uobject}.
Correspondingly, divergent branches can be combined (or merged), for example $v_2$ and $v_4$ are merged into
$v_5$. This abstraction can be viewed as an extension of key-value, in which the application can treat each
update as a ({\em key, version, value}) tuple, and retrieve existing values via composite keys.

\subsubsection{Versions}
In \ustore, version numbers not only serve as unique identifier of \unode\ objects, they also play a key role
in security and load balancing. More specifically, \ustore\ ensures three properties of version numbers:
\begin{enumerate}
\item {\bf Unique:} \unode\ objects have distinct version numbers. 
\item {\bf Verifiable:} Version numbers can be used to prove integrity of the retrieved objects. 
\item {\bf Locality-aware load balancing:} \unode\ objects can be partitioned over multiple nodes, and objects of the same
branch have high probabilities to be stored in the same node. 
\end{enumerate}
Essentially, a \ustore\ version is a 3-field tuple:
\[
v=(v.h, v.l, v.n)
\]
\noindent where $v.h$ is the cryptographic hash tied to the object content, $v.l$ is the depth (from
root of the DAG), and $v.n$ is a random value generated by a storage server. The root version is defined as $v_{\text{\em NULL}} = ({\text{\em NULL, NULL, NULL}})
$. We explain how \ustore\ achieves its three properties later in this section.

\subsubsection{APIs}
To write a new value to a key, the application needs to specify the base version from which the new value is
derived.
Optionally, it can specify whether it wants to compress the value,
which we discuss later in Section~\ref{subsec:storage}.
\begin{align*}
&{\text{\tt Put(key, base\_version, value, compress)}}\\
& \qquad \qquad \to {\text{\tt version}}
\end{align*}
\noindent When receiving the request, \ustore\ creates a \unode\ object for the value and connects to the base
object. A new version number is returned so that the application can retrieve this value later. The most
important advantage of this operation is high throughput, as no requests are blocked or aborted.  Since
requests with the same {\tt base\_version} will result in multiple branches, no locks are required and thus
many requests can be served at
the same time. 

To retrieve a value, the application supplies both a key and a version number in order to
identify the unique data version. 
\[
{\text{\tt Get(key, version)}} \to {\text{\tt value}}
\]
\noindent \ustore\ first locates the target \uobject\ based on the key, then it returns the corresponding \unode\ value
for the given version. Note that this operation is also non-blocking, as existing versions are immutable and
available for reading all the time. \ustore\ provides no APIs for getting all the latest versions of a key,
i.e., the versions without any successors.  This is our design choice to minimize the overhead incurred by
consistency mechanisms when there are concurrent updates. Nevertheless, we provide an option for the application
to receive notifications of updates via a publish/subscribe channel. 

To merge two branches, we rely on the applications to specify merge semantics (discussed later), and only
pass the consequent merged value to the storage.  
\begin{align*}
& {\text{\tt Merge(key, version\_1, version\_2, merged\_value)}}\\
& \qquad \qquad \to {\text{\tt version}}
\end{align*}
\noindent \ustore\ handles this operation similarly to a {\tt Put} operation, but creating two connections to the two
base \unode\ objects. Data immutability is not affected by {\tt Merge}, since the operation creates a new version for
the merged value. 

To track previous versions or values of a given \unode, the application can invoke:
\begin{align*}
& {\text {\tt GetPreviousVersion/Value(key, version)}} \\ 
& \qquad \qquad \to \{\text{\tt versions/values}\}
\end{align*}
\noindent which returns a single version/value of the current \unode\ as derived from a put operation, or
two versions/values if it is derived from a merge operation. For example, in Figure~\ref{fig:uobject}, given
$v_8$, this operation returns $v_7$, and given $v_5$ it returns $\{v_2, v_4\}$. When receiving this request,
\ustore\ first locates the current \unode\ using the given version, then follows the backward connections to
fetch the previous objects. By invoking this operation repeatedly, the application can trace the full history
of the \uobject. 

To avoid calling {\tt GetPreviousVersion/Value} many times, the application can use a batch operation that
retrieves up to $m$ previous \unode s. 
\begin{align*}
& {\text {\tt GetKPreviousVersion/Value(key, version, m)}} \\
& \qquad \qquad \to \{\text{\tt versions/values}\}
\end{align*}
\noindent \ustore\ first locates the current \unode\ object, then recursively fetches the previous object,
stopping when one of two following conditions occurs: when it reaches $m$ hops away from the original
requested object, or when it encounters a merged object. When the operation returns $m$ objects with versions
$\langle v_1,v_2,..,v_m \rangle$, it means that $v_i$ is the predecessor of $v_{i-1}$, $v_1$ is the predecessor of {\tt version}, and there
are no branches in between. When it returns $m' < m$ objects, it means there is a branch at $v_{m'}$. Using
this information, the application can determine which branch to follow next. For example, {\tt
GetKPreviousVersion}$(k,v_8,3)$ returns $\langle v_7, v_2, v_1 \rangle$, whereas {\tt GetKPreviousVersion}$(k,v_9,3)$ returns
$\langle v_6, v_5 \rangle$.

\subsection{Physical Layer}
We now describe how \ustore\ implements the abstraction above at the physical layer. Our current design assumes
all data is kept in memory, and support for migration to secondary storages is part of future work.  


\subsubsection{\uobject\ storage and indexing}
\label{subsec:storage}
There are two strategies to materialize \uobject's content: complete or incremental.
In the former, \unode\ objects are stored in their entirety. 
In the latter, each object contains only the compressed data (e.g, the delta, which is the
{\em diff} with its previous version), thus significantly reducing the storage consumption for a large \uobject.
However, there is a trade-off between the storage consumption and computation cost to reconstruct the objects. Thus, for
the incremental strategy, in order to avoid traversing a long path to get a complete value, \ustore\ allows the
application to specify whether it wants to compress or not at each step.
In addition, the application can register its specific
\textit{compress()} and \textit{decompress()} functions
based on their own data characteristics,
which will be used by \ustore\ during the compression and de-compression processes.
As a result, \ustore\ is flexible enough to achieve a variety of compression strategies.
In {\ustore}, the value is compressed based on its previous versions
only if its previous version is local and is uncompressed.
Otherwise, it will be based on its nearest uncompressed ancestor within the same node.
The locality-aware partitioning scheme (discuss later),
makes it highly possible that the previous version is located in the same node.
We do not choose to compress the data based on a compressed one,
because this will increase both the compression and de-compression cost.

As a \unode\ is uniquely identified by its key and version, we adopt a simple
hash-based indexing to quickly locate the object with $O(1)$ complexity. 
The {\tt Put}$(k,v_p,o)$ generates a new \unode\ version number as follows: 
  \begin{equation}
  v = (v.h, v.l, v.n) = (H(k||v_p.h||v_p.l||o), v_p.l+1, \eta)
  \label{eq:versions}
  \end{equation}
where $H$ is a cryptographic hash function and $\eta$ a random value generated by the storage node. The version
generated by {\tt Merge}$(k, v_{1}, v_{2}, o)$ is similar, except that $v.l = \textit{max}(v_{1}.l,
v_{2}.l)+1$. This means a merged object will be on the longest branch of its two ancestors, for example in
Figure~\ref{fig:uobject}, $v_{5}.l=3$.  $H$ ensures that version numbers are verifiable because both the content
and link to the previous version are used to compute $v.h$. It also ensures uniqueness: for any two inserts
$v_1 \leftarrow \text{\tt Put}(k,v_p,o)$, $v_2 \leftarrow \text{\tt Put}(k',v_p',o')$ such that $(k,v_p,o)
\neq (k',v_p',o')$, then $v_1 \neq v_2$.  


\subsubsection{Locality-aware partitioning}
\label{subsub:part}
\uobject\ granularities and volumes may vary considerably, for instance from large numbers of small objects
as in a database application, to small numbers of large objects as in a Git-like application.  In order to
scale out to many nodes, we need to distribute  {\uobject}s evenly to multiple storage nodes,
which can be achieved by hashing the key. The challenge, however, is in partitioning a single, large \uobject\ to
multiple nodes, as shown in Figure~\ref{fig:uobject}. 

One approach to achieve load balancing is to distribute \unode\ objects based on their versions using consistent hashing.
Specifically, a version $v$ is first hashed to identify the storage node, then it is used to index into the node's local hash
table. However, this approach fails to preserve locality: if $v_2$ is derived from $v_1$, the
probability of both being in the same node is the same as that of any two random versions.
Figure~\ref{fig:uobject} shows an example of locality-preserving partitioning schemes which distributes
related objects together onto the same node. Note that locality hashing schemes are not useful in our case,
because the way we compute version numbers using a cryptographic hash function (Eq.~\ref{eq:versions})
destroys the relationship between two related versions. Instead, \ustore\ achieves this property by generating
versions using additional inputs from the storage nodes. 

For each \uobject, each server reserves a memory region $t$ for storing its \unode s. The application can adjust $t$ dynamically
to better suit its storage requirements\footnote{via an out-of-band protocol}. For a write request, i.e. {\tt Put}$(k,v_p,o$) where $v_p =
(v_p.h, v_p.l, v_p.n)$, the application uses $(k||v_p.n)$ to locate a storage node $S$ via consistent hashing, and
routes the request to $S$. The node $S$ handles the request as follows: 
\begin{itemize}
\item If its reserved memory region for the \uobject\ is not full, it stores the object and returns a
\texttt{SUCCESS} status, together with a new version $v = (H(k||v_p.h||v_p.l||o), v_p.l+1, v_p.n)$. 
\item When the reserved region is full, it generates a random value $\eta$ such that $(k||\eta)$ maps to a
different region in the consistent hashing space. It returns a \texttt{REDIRECT} status and $\eta$ to the
application. 
\end{itemize}
When the application receives a \texttt{REDIRECT}, it repeats the write request, but using
$(k||\eta)$ for consistent hashing, until it gets a \texttt{SUCCESS} response. The application can ask the
server to increase $t$ after a pre-defined number of redirected request. In one extreme, for example, the
application may want to ensure all the write operations where $v_p = v_{NULL}$ (root versions of top-level branches)
result in the data being stored in the same node. In this case, when \texttt{REDIRECT} is returned, it asks the server to
increase $t$ immediately instead of following the redirection.

This protocol ensures object locality, because an write operation based on version $v_p$ implies the new object has
derivation relation with $v_p$, and its first priority of storage node is the same as that of $v_p$. Only when
the node exhausts its capacity is the object {\em spilled over} to another node. Unlike the common
load-balancing approach in distributed hash tables which uses redirection pointers, \ustore\ returns a
new mapping in the form of $\eta$, thus future requests are routed directly to the new node. In particular, a
read operation, i.e. {\tt Get}$(k,v)$, proceeds by forwarding the request to the storage node identified via
consistent hashing of $(k||v.n)$. The node finds the object indexed by $v$ and returns the object. By being
able to adjust $t$, the application has full control of how to balance the workload.

\subsubsection{Caching and replication}
\label{subsub:caching}
The locality afforded by \ustore's partitioning does not help when objects are in different nodes. For example, given
the objects in Figure~\ref{fig:uobject}, the request {\tt GetKPreviousValue}$(k,v_5)$ incurs 3 network requests sent to
3 different servers. \ustore\ provides a caching layer, that is each server maintains a cache of remote objects fetched
during scan operations. In our example, $v_2$ and $v_4$ are cached at the same node of $v_5$, thus the next request for
immediate predecessors of $v_5$ can be returned right away. By exploiting temporal locality of scanning operations, the
server can answer requests more quickly. The cache is simple to maintain, since there is no cache coherence problems
with immutable objects. For each \unode\ object, the server caches up to 2 remote predecessors, because the benefit of
caching diminishes with increased number of cached predecessors. 

\begin{figure}
\centering
\includegraphics[scale=0.35]{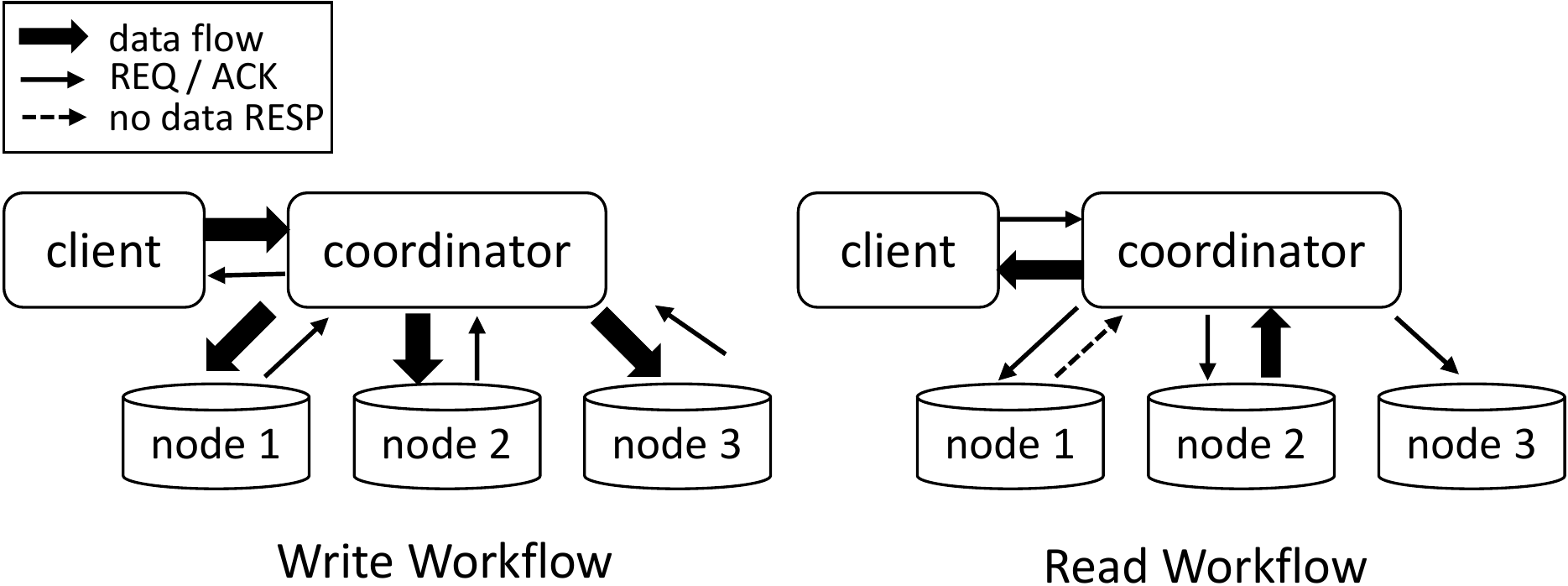}
\caption{Workflow of write and read request}
\label{fig:workflow}
\end{figure}

\ustore\ replicates \unode\ objects to achieve scalability and fault tolerance. Neighboring nodes in the
consistent hashing space are replicas of each other. Replication is controlled by two parameters: number of
replicas $N$ and a write set size $W$. Figure~\ref{fig:workflow} illustrates the workflow for write and read
requests (for simplicity, we assume that all inserts return \texttt{SUCCESS}). During a write
operation, the {\tt Put} request is sent to all the replicas in parallel, and it returns once there are
acknowledgements from $W$ servers. Note that $W$ guarantees that the data is available when there are no more
than $W$ failures during the write operation. For a {\tt Get} operation, the request is sent to the replicas
in turn until the requested data is returned. 

It can be seen that $W$ balances the cost of read and write requests: larger $W$ increases write latency,
but may reduce read latency with high probability.  On the other hand, larger $W$ means less availability for
write operations which only succeed when there are fewer than $(N-W)$ failures, but increased availability for read,
as there are more replicas with the data. We note that there is similarity between \ustore's replication
scheme and Dynamo's~\cite{dynamodb} which is configurable by two parameters $W$ and $R$. \ustore\ does not
require specifying $R$ (or in fact, in \ustore $R=1$), because there is no read inconsistency in
our system, thanks to the data being immutable. 

\subsection{View Layer}
This layer enhances the immutability semantics of the layers below with fine-grained access control, data
security. It increases the system's flexibility by adding a notification service and an application hook for
pushed-down merge semantics. 
\newline
\subsubsection{Access control and data security} 
\ustore\ enables fine-grained access control at \uobject\ level, allowing users to specify policies concerning specific
\unode\ objects. The current design supports only read policies, and it assumes that storage servers are trusted, that
users are authenticated (Section~\ref{sec:discussion} discusses the challenges in supporting more expressive policies
with stronger trust models). By default, \uobject s are readable by all users. For a private \uobject s, \ustore\
maintains a {\em shadow} object, called, \acuobject, with the same key. The \acuobject\ contains policies of 
the form $\langle \texttt{userId, list of versions}\rangle$, indicating which users can read which versions of the
associated \uobject. 

Only the owner of the \uobject\ can update its corresponding \acuobject, using the extended {\tt Put} API. The
\acuobject\ is stored at the same server with the versions being protected. When receiving a read request for some local
versions, the server scans the local \acuobject\ for a policy granting access to the versions, before returning them to
the user. Each new policy results in at most $N$ \acuobject s where $N$ is the total number of servers storing the
associated \uobject. This overhead can be mitigated by batching multiple policy into the same update. We also note that
the number of policy updates in \ustore\ is smaller than in other systems with mutable states, because there is no need
for revocation (once read access is granted to a version, it is considered permanent).  



\ustore\ protects integrity of both the data and of the version history against untrusted servers. Specifically, the
version number is computed via a cryptographic hash function over both the data content and previous version number 
(Eq.~\ref{eq:versions}), thus it is not possible for the untrusted server to return different data (for {\tt Get}
operations) or different predecessors (for {\tt GetPreviousVersion/Value} operations). We consider two update operations
with the same data and the same base version to be duplicate, and therefore it is safe to return the same version. For
stronger guarantees, each server can also generate signatures for successful {\tt Put} operations that can be used as
proof of data storage. 


\subsubsection{Consistency model}
Many distributed storage systems adopt the eventual consistency model: they allow reads to see stale values.
However, reasoning about the returned value of a read is difficult in this model, because it depends on a multitude of
factors: the concurrency model, the read-repair protocols, etc. In \ustore, there are no stale reads, since
every {\tt Get} operation is specific to a version which is unchanged.  There are only two possible outputs of
a read operation: the correct value, or an error. The error indicates that the version has not been propagated
to this replica, thus the request should be retried later.  This semantics is clean and simple, making it
easy to reason about the system's states and correctness. Write operations in \ustore\ are highly efficient
for two reasons. First, once a \unode\ is written, it does not have to be sent immediately to other replicas.
  Second, concurrent writes to the same object in \ustore\ have no order, therefore they require no locks and
  can be executed in parallel. 

{\bf Merge semantics.} Applications built on \ustore\ must explicitly deal with conflicting branches caused by
concurrent writes.  \ustore\ supports branch reconciliation via a function which merges two branches together.
Specifically, the function \texttt{Merge}$(v_1,v_2,f_m) \to \{(o),\bot\}$ takes as input two version $v_1$, $v_2$ belonging
to two branches, a user-defined function $f_m$, and generates a merged value defined by $f_m$
(if successful). Our current design uses 3-way merge strategy, although we note that there exists several
alternatives. This merge function first finds the closest ancestor, say $v_0$, to both $v_1$ and $v_2$.
$v_0$ is where the two branch containing $v_1$, $v_2$ forked.  Next, it invokes $f_m(v_0,v_1,v_2)$ to perform
3-way merge, which returns a value $o$ if successful.  Finally, it calls the lower-layer function {\tt Merge} to
write a new version to \ustore. 

We observe that different applications may follow different logics when merging branches, and user's intervention maybe
needed because the conflict cannot be resolved automatically. For example, in Ficus~\cite{ficus}, two versions adding two
files to the same directory can be merged by appending one file after another. In Git, versions that modified two
different lines can be merged by incorporating both changes, but if they modified the same line the merge
should fail. \ustore\ allows an application to define its own function $f_m(.)$ and use it when initializing the store.
There is a number of pre-built functions in \ustore, such as append, aggregation and choose-one. 

\subsubsection{Notification service}
Recall that \ustore\ does not provide APIs for retrieving the latest \unode\ objects. The main reason for
omitting these APIs is to keep the storage semantics simple and clean. Under replication and failure,
reasoning about the latest version is difficult, since the APIs may return different results for the same
two requests. However, the need to track latest versions is essential to many applications, but we also note 
that many applications do not require real-time notification, hence they are tolerant of notification delays.
One example is the collaborative analytics application in which collaborators need to be aware of each other's
updates in a timely manner, but not necessarily in real time, in order to avoid repeating work and complex conflicts.

To enable version tracking, \ustore\ provides a notification service to which applications can subscribe. The
service is essentially a publish/subscribe system in which the storage servers are the publishers and applications are the
subscribers. It maintains pairs of events and application IDs, and is responsible for routing the
events to the appropriate applications. When an application wishes to be informed of new versions of a \uobject, it
invokes {\em register(.) } function with its ID and the key of interest. When there is an update to the
\uobject, the storage server creates a new version and invokes {\em publish(.)}. The service receives the new
version and routes it to the registered subscribers. Our current design uses Zookeeper for this service, but we
are adapting the design of Thialfi~\cite{thialfi} to make the service more scalable.

\section{\ustore\ Applications}
\label{sec:newapp}
In this section, we present our implementations of four applications on top of \ustore. They are a mix of
established and emerging applications that exploit \ustore's core APIs to achieve high performance while reducing
development effort. 

\subsection{Git} 

There are four main types of data structures in Git: \emph{Blob} which contains unstructured data (e.g., text or binary
data); \emph{Tree} which contains references to other blobs and trees; \emph{Commit} which contains meta data for a
commit, i.e., commit message, root tree reference and previous commit reference; and \emph{Tag} which contains tag name
and commit reference. These data types are managed in a key-value store as records, and object references (i.e., keys
of an object record) are explicitly recorded in the content.  As a result, to extract references, the whole record has
to be fetched and de-serialized. In \ustore, we can easily separate history-based references (e.g., previous commit of a
commit) from content-based references (e.g., blobs in a tree) since the storage's version tracking naturally supports
history-based references. We discuss here a simple version of Git implemented in \ustore, providing the same properties
as the existing implementation. We refer readers to the Appendix for an extension of this design that provides richer
functionalities, such as file-level history tracking. 

The original Git implementation supports content-addressable storage by identifying an object by the cryptographic hash
of its content. In \ustore, version numbers can readily be used to uniquely identify objects. We maintain one \uobject\ for
each data type $T$ which then manages all objects of that type.  Fetching an object with a hash content $h$ can be done via
{\tt Get}$(T, h)$. Similarly, to commit a new version with content $o$, we use {\tt Put}$(T, \text{NULL}, o)$, which
returns a deterministic and unique version. Since the version number can be pre-computed,
we can check if the version exists before committing. Note that when writing a {\em Commit} object to \ustore,
the previous version is tracked implicitly, enabling fast traversal of the commit history. Other Git commands, such as
checkout, branch and merge can be implemented directly on top of these two fetch and commit primitives.   



One benefit of using \ustore\ is the ability to separate content and history references, making it more efficient to
implement commands like \texttt{git log}.  Furthermore, we no longer need to fetch the whole repository to check out a
specific version, which is inefficient for repositories with long histories.  Another benefit comes from \ustore's
flexibility to support customized compression functions which can be more effective than the default \emph{zlib}
function. Our implementation totals $289$ lines of C++ code. As a reference, the Git codebase adds up to over
$1.8$ million lines of C code (but it supports many more features than our \ustore\ based implementation).

\subsection{Collaborative Data Science}
It is becoming increasingly common for a large group of scientists to work on a shared dataset but with different
analysis goals~\cite{adam_sigmod15,datahub}.  For example, on a dataset of customer purchasing
records, some scientists may focus on customer behavior analysis, some on using it to improve inventory
management. At the same time, the data may be continually cleaned and enhanced by other scientists.  As the
scientists simultaneously work on the different versions or branches of the same dataset, there is a need for
a storage system with versioning and branching capabilities.  Decibel~\cite{decibel} is one of such
systems that supports relational data model.  We implement an application on \ustore\ for the
same collaborative workflows supported in Decibel. 

Like Decibel, we provide two storage strategies for versioned relational databases, namely {\em Tuple-First} and
{\em Version-First}. In the former, we treat tuples from different tables and versions the same. Specifically,
a tuple is stored as a \uobject, where the key is the tuple's primary key and the different versions of the
tuple correspond to the \unode\ objects. To realize the relational model, we use another type of \uobject\ to
maintain the tuples' membership to tables and versions, where the key is the table name and version, and each
\unode\ stores a bitmap index to track whether a tuple is present in the versioned table.  In the
Version-First strategy, tuples from one versioned table are stored in a single \uobject\ object. We support
large numbers of tuples by storing them with two-level paging. More specifically, in the first level, the
\uobject's key is the table name and version, while its value contains a set of keys of the second-level
\uobject s which actually store the tuples. 

There are two advantages in our \ustore-based implementation compared to Decibel. First, \ustore\ stores
tuples in memory instead of on disk, thus the operations are faster. Second, it can be scaled out easily to
support large datasets, whereas Decibel is currently restricted to a single node. Our implementation amounts
to $640$ lines of C$++$ code (as a reference, the Decibel's codebase adds up to over $32$K lines of Java code).

\subsection{Transaction Management}

TARDiS~\cite{tardis} is a branch-and-merge approach to weakly consistent storage systems.  It maintains a data structure
called {\em state DAG} to keep track of the database states as well as the availability of data versions to any
transaction.  Unlike Git which creates branches explicitly on demand, TARDiS generates branches implicitly upon conflicts
of data accesses.  By doing so, TARDiS can keep track of all conflicting data accesses in branches. Like \ustore, it
enables flexible conflict resolution by allowing the high-level application to resolve the conflicts based on its own
logic.

In TARDiS, a transaction $T$ issued by client $C$ starts by searching the state DAG for a valid state that it can read
from.  A state is valid when it is both consistent and compatible to $C$'s previous commits. It could be a state that is
previously created by $C$ or whose ancestor state is created by $C$.  Once a read state is selected, denoted as $S_{T}$,
the transaction can perform read operations by referring to $S_{T}$.  Specifically, when reading an object $O_{r}$, it
checks all the versions of $O_{r}$ in the storage starting from the latest version, and greedily finds the version
(identified by the corresponding state which created it) that is compatible with $S_{T}$. Here, compatibility means that
the version $T$ is reading must be in the same branch with $S_{T}$.  When writing an object $O_{w}$, the transaction 
creates a new version of $O_{w}$ based on the transaction identifier.  When committing, it checks whether $S_{T}$ is
still valid.  As long as $S_{T}$ is valid, $T$ will eventually commit (e.g., commit after the $S_{T}$ in the state DAG).
Because other concurrent transactions may not have conflicts with $T$ in terms of updates, it {\em ripples down}
from $S_{T}$ to find and commit after the deepest compatible state. 

The structure of state DAG in TARDiS can be mapped directly to \ustore.  In fact, we implement TARDiS in \ustore\ by
simply using \uobject\ to store the states. This implementation, referred to as TARDiS+\ustore, leverages \ustore's
efficient version scan operation to carry out backward search from the latest DAG states.  When data
accesses are skewed, there is a high probability that data read by a transaction is updated in a recent state.
Therefore, searching for data versions by backtracking (as in TARDiS+\ustore) is more efficient than by
scanning the topologically sorted version list for each data item (as in TARDiS). Our
implementation adds up to $1068$ lines of C++ code (there is no open source version of TARDiS, so we implement
both systems from scratch).

\subsection{Blockchain}
\label{sec:app_blockchain}

\begin{figure}
\centering
\includegraphics[scale=0.38]{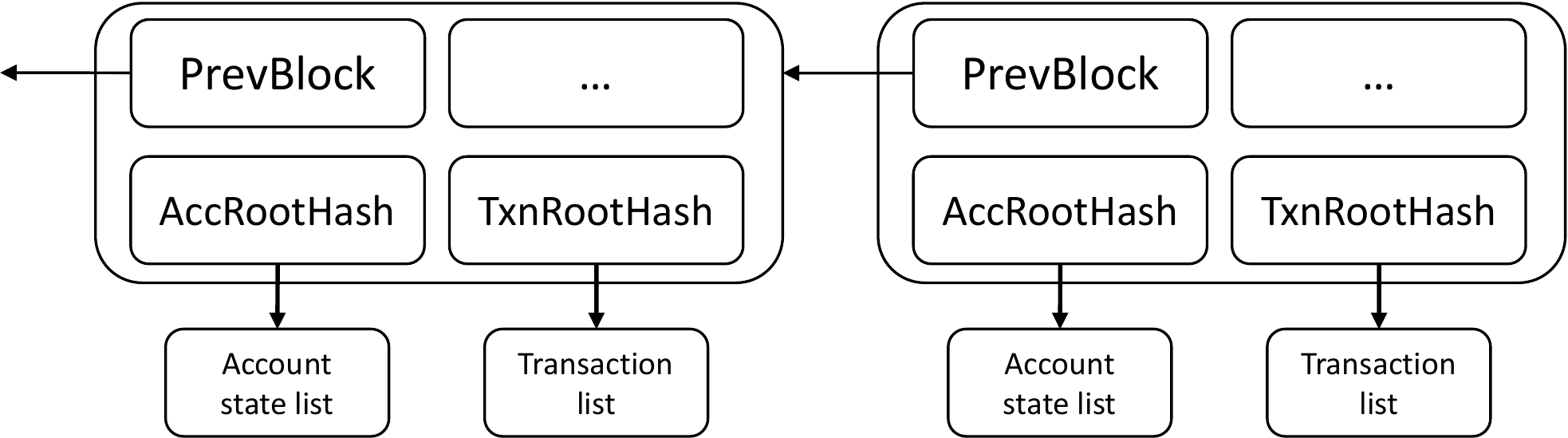}
\caption{Blockchain data structure, transaction and account list are stored beyond block layer.}
\label{fig:blockchain}
\end{figure}

A blockchain is a decentralized shared ledger distributed among all participants of the system. Data
immutability and transparency are two key properties that fuel the recent rise of blockchains. Figure
\ref{fig:blockchain} illustrates a typical blockchain data structure, where the data is packed into a block
that is linked to the previous block, all the way back to the first (genesis) block, through cryptographic
hash pointers. If the data is tampered in any block, all hash pointers in the subsequent blocks become
invalid.  In a blockchain network, the
participants agree on a total order of transactions, i.e. a unique evolution history of the system states. In
some blockchain systems, like Bitcoin \cite{bitcoin}, the blocks keep no account state (e.g., balance),
but only the unspent crypto-currencies (or coins). In other systems such as Ethereum
\cite{ethereum}, the account states are stored explicitly in the blocks as shown in Figure
\ref{fig:blockchain}.

We implement a blockchain data structure similar to Ethereum using {\ustore}, in which we maintain the account states inside the blocks. There are
two layers of data in our design, namely a block layer and an account layer. In the block layer, each block
contains metadata, such as the proposer of this block, the root hash of account list, etc. It is stored as a
\unode\ object whose key is the same for every block. When we append a new
block (using {\tt Put}), we use the version of its preceding block as its base version. This way, \unode's
version numbers become the hash pointers of the blockchain. In the account layer, each account object, stored
as a \unode, keeps track of the account balance, where the \unode's key is the account address. 

Existing blockchain systems still lack an efficient and scalable data management component, which also helps
enhance the security and robustness of the blockchain. For instance, the recent DDoS attack on Ethereum 
is attributed to inefficiency in fetching state information from disk. A new blockchain built on \ustore\ can
benefit directly from the scalability and efficiency of the storage. Moreover, by exploiting \ustore's
versioning capabilities, the new blockchain system can support efficient analytical queries which are useful for gaining insights from the data in the blockchain.
We implemented the blockchain logics on \ustore\ using 1,231 lines
of code in C++. As a reference, the popular Ethereum client (Geth) comprises 539,584 lines of Go code. 


\section{Evaluation}
\label{sec:evaluation}
In this section, we report the performance of \ustore\ and of the four applications discussed above. We first
evaluate \ustore's data access operations (read, write and version scan) and compare them against Redis.  The results
show that \ustore\ achieves comparable performance with Redis for basic read and write operations.  
By exploiting locality-aware partitioning, \ustore\ achieves $40$x lower latency than Redis for scan operations.  We
also examine the performance trade-off against availability and compression strategies, which can be controlled by the
applications.  Next, we evaluate four \ustore-based applications against systems supporting the same operations. The
Git-like application improves commit and checkout latency by up to $240$x and $4000$x respectively, thanks to
data being stored in memory and the simplicity of the checkout operation. The collaborative data science application 
achieves up to $10$x better latency in 3 out of 4 queries, due to the in-memory design.  The transaction management
application reduces reduces the number of states accessed by up to $80$x for skewed workloads by leveraging the
efficient scan operations.  Finally, for blockchain application, \ustore's datastructure matches well with blockchain
data, and the system's efficient data access operations account for up to $400$x lower latency in 4 out of 5 queries.  

All experiments were conducted in a 20-node, RDMA-enabled cluster. Each node is equipped with a E5-1650 3.5GHz CPU,
32GB RAM, 2TB hard drive, running Ubuntu 14.04 Trusty. There are two network interfaces per node: one 1Gb Ethernet, and
one Mellanox 40Gb Infiniband.
\newline

\subsection{Microbenchmark}
\begin{figure}
\centering
\includegraphics[width=0.35\textwidth]{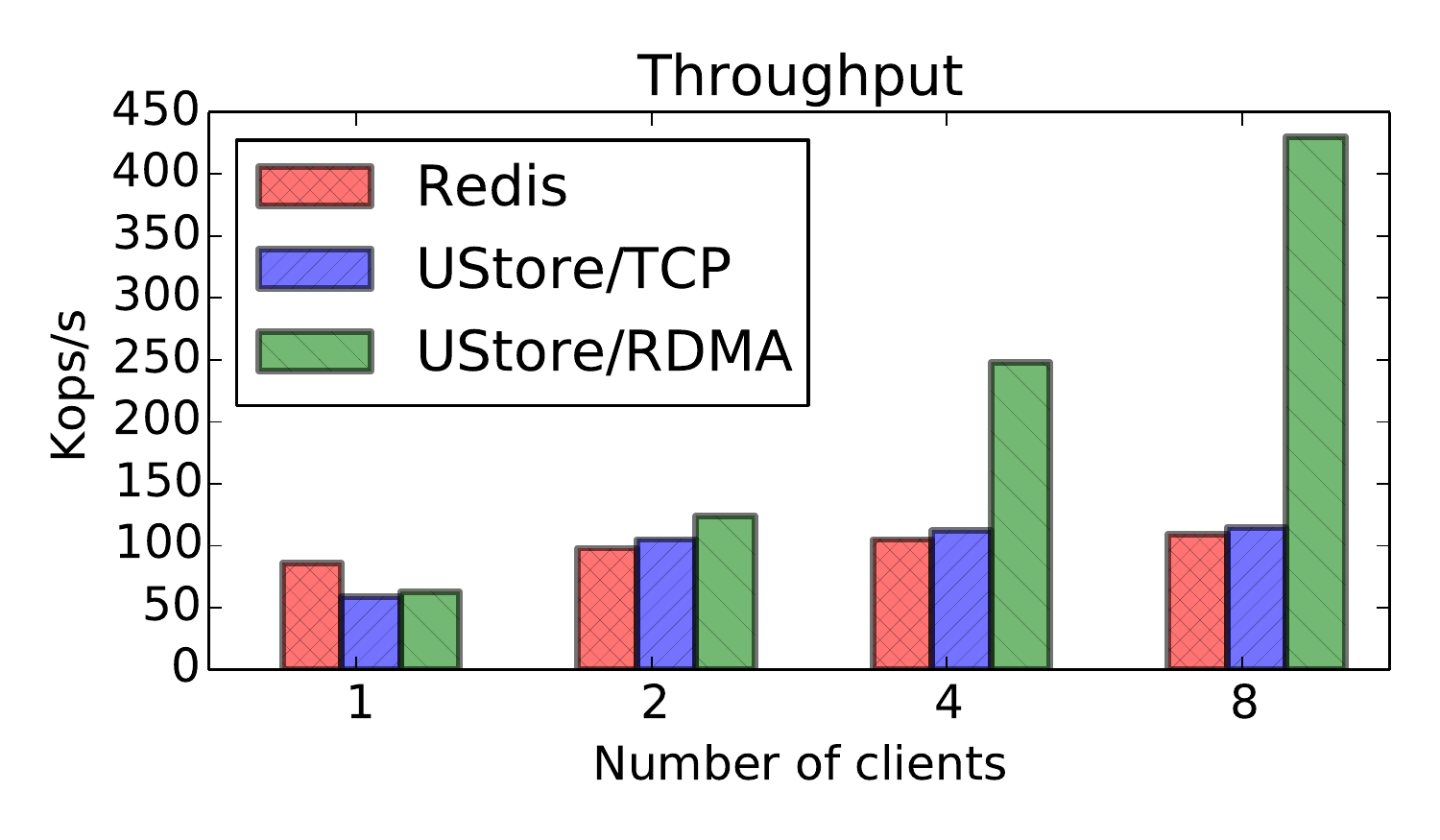}
\caption{Throughput vs. Redis}
\label{fig:tp}
\end{figure}
\begin{figure}
\centering
\includegraphics[width=0.35\textwidth]{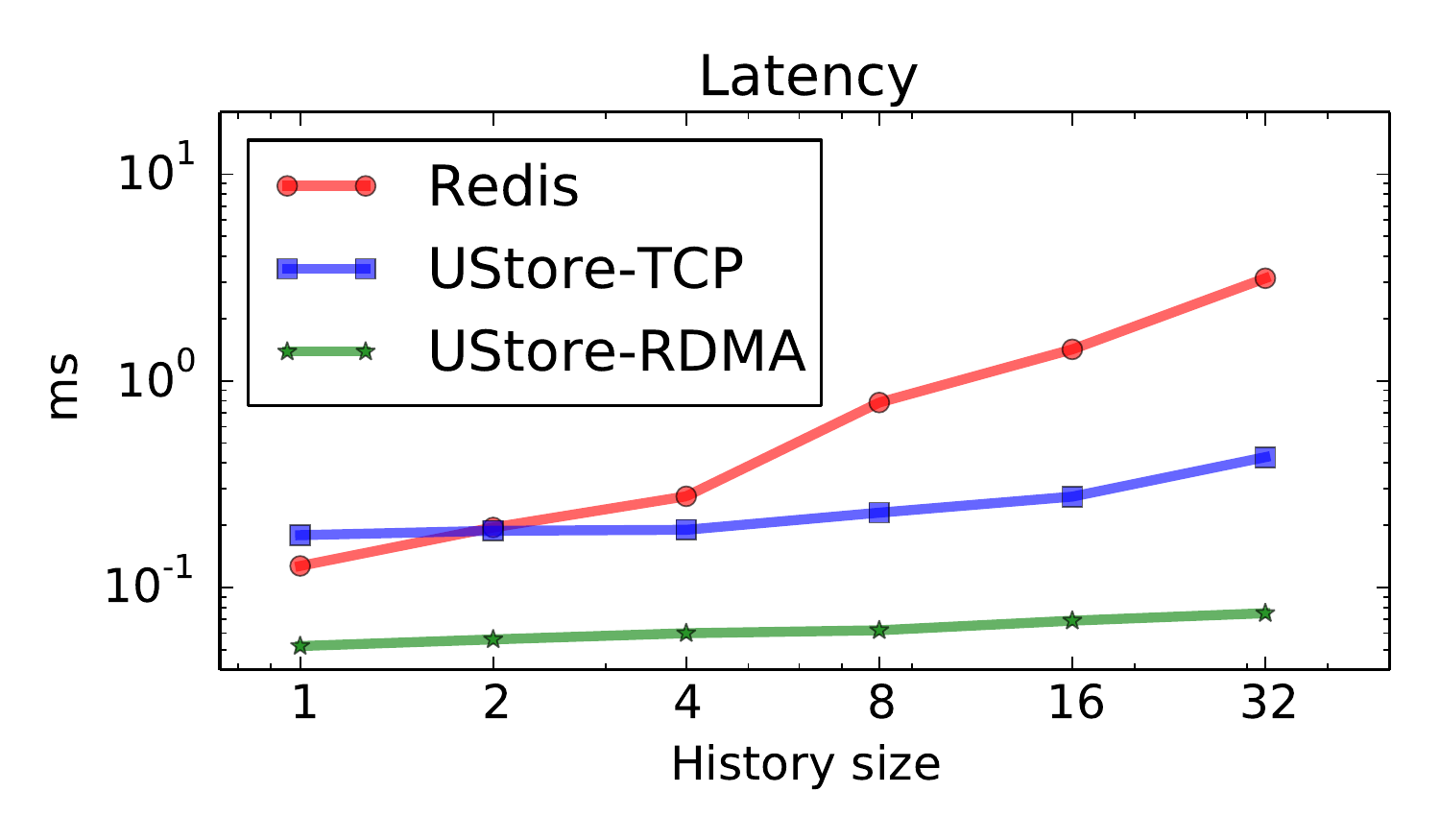}
\caption{Branch scan queries}
\label{fig:scan}
\end{figure}
{\bf Basic operations.} We ran the YCSB benchmark to measure \ustore's read and write operations and compare them
against Redis ($v3.2.5$ release).  To support pure key-value workloads (without versioning), we replace \ustore's
default hash function with another that ignores version numbers when locating the objects. Figure~\ref{fig:tp} shows the
throughput for a varying number of clients over the workload with $50/50$ read and write ratio, each client using $64$
threads. Without RDMA, \ustore's performance is comparable to Redis's. With a single client, Redis achieves higher
throughput, at $86$K operations per second (Kops), than \ustore\ does at $60$ Kops. This is due coarse-grained locks
implemented at \ustore\ clients (future optimization will likely improve the current client throughput). However, with
more clients, Redis server becomes saturated and \ustore\ performs slightly better. The performance gain is due to
multi-threaded implementation of \ustore, as opposed to Redis' single-threaded implementation. Using RDMA, \ustore\
achieves much higher throughputs, which is as expected because of the high bandwidth and efficient networks. In
particular, the RDMA version requires 8 clients to saturate it at $430$ Kops, whereas Redis is saturated by 4 clients at
$110$ Kops. We note that this gap could be eliminated when Redis includes native support for RDMA. 

{\bf Scan operations.} Next, we considered version scan operations which are common in many applications. Given a
version $v$ and value $m$, the query returns $m$ predecessors of $v$, assuming a linear version history. This is
supported directly in \ustore\ by the {\tt GetKPreviousValue/Version} APIs. We implemented this operation in Redis for
comparison, in which we embedded the previous version number into the value of the current version.
Figure~\ref{fig:scan} illustrates the differences in latency between \ustore\ and Redis. With increasing $m$, Redis
incurs an overhead growing linearly with $m$, since it must issue $m$ sequential requests to the server. On the other
hand, the number of requests in \ustore\ is constant since the server can send up to $m$ versions back in one response.
Since each request involves a network round-trip, the query latency at $m=32$ is much lower in \ustore, at $0.4$ms (and
$0.07$ms with RDMA), than in Redis (over $3$ms). 

{\bf Performance trade-offs.} We benchmarked \ustore's various performance trade-offs by examining its performance with
different values of $W$ and different compression strategies. We briefly explain the findings here, and refer readers to
the Appendix for the detailed results and analysis. We observed that increasing $W$ leads to slower writes, but it has
no impact on read latency. Furthermore, our default compression strategy outperforms a random compression strategy, and
achieves a good balance in terms of throughputs and memory consumptions as compared to a no-compression strategy. 

\subsection{Git}
\begin{figure}
\centering
\includegraphics[width=0.35\textwidth]{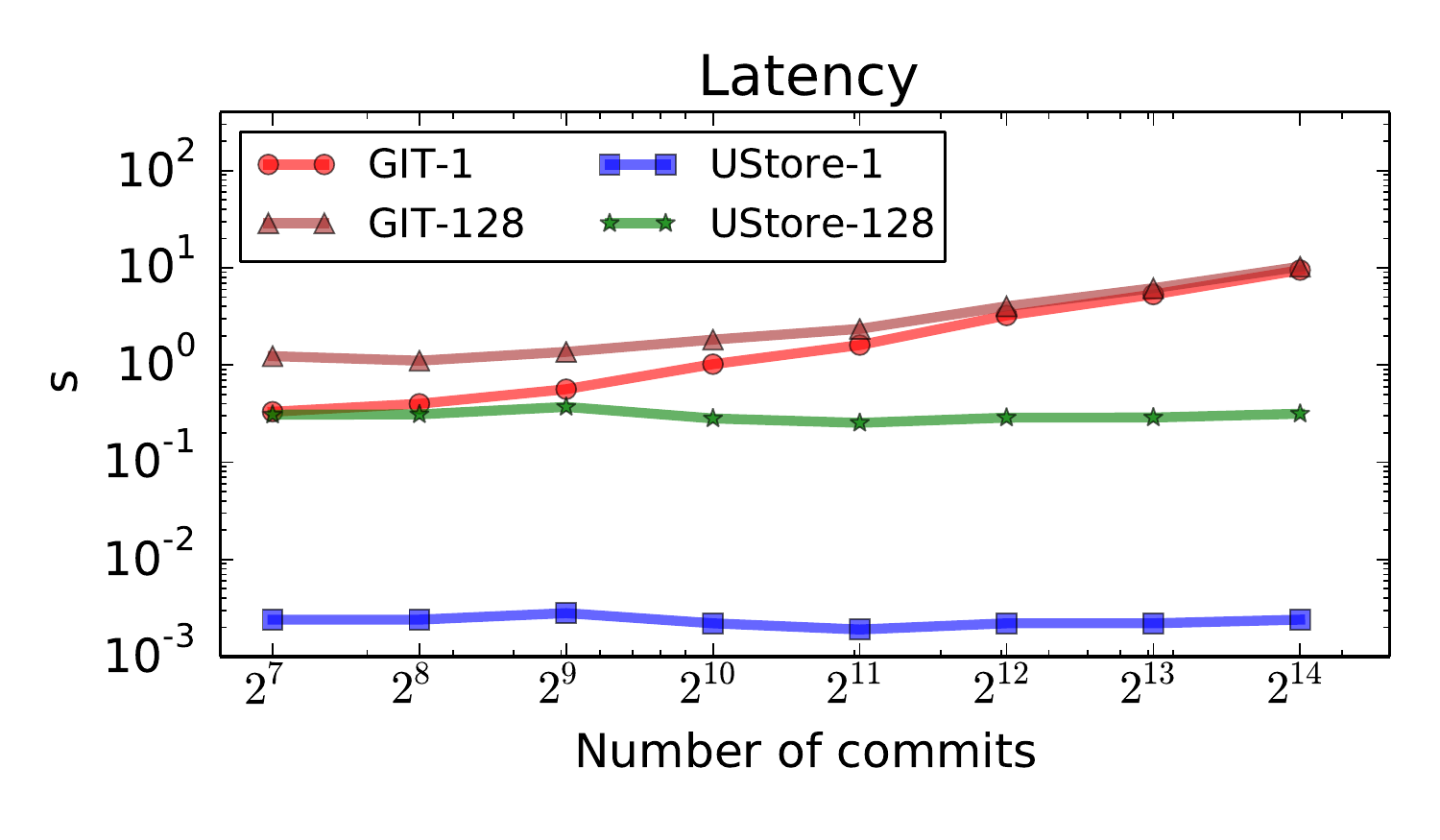}
\caption{Checkout latency in origin Git vs. in \ustore. Git-X and \ustore-X mean there are $X$ different
checkout operations. }
\label{fig:git_checkout}
\end{figure}

We evaluated our \ustore-based Git implementation (referred to as \ustore) against the original Git in two
data-versioning related operations, i.e. checkout and commit.  We measured both implementations in terms of operation
latency and storage consumption. First, we generated a synthetic workload containing varying numbers of data versions.
Each version consists of a single fixed-size ($32$KB) file with random printable characters.  For each commit operation,
we overwrite the file with content of the next version, and commit it into Git or \ustore. For Git, we use a practical
work-flow for commits, in which each commit is first saved into a local repository ({\tt git commit}), and then pushed
to a remote server ({\tt git push}) running GitLab.  For \ustore, we set up one server and have the client sending
requests from a remote node.  

Figure~\ref{fig:git_checkout} shows the latency for checkout operations with varying repository sizes (number of
commits). We refer readers to the Appendix for the results of commit operations and of the storage cost.  For
single-version checkout operations, Git is significantly slower than \ustore\ ($4000$x), because the former requires the client to fetch the
entire history even when only a single version is needed. The checkout latency in Git also increases with longer
history, whereas \ustore's latency remains constant, because the latter fetches only one version. For the multiple-version
checkout operations, we observe similar, but smaller gap. In this case, \ustore's latency increases linearly with the
number of checkouts because each operation is independent, whereas in Git the initial cost of fetching the entire
history is amortized over subsequent checkouts which are done locally. 

To compare the cost of commit operations in \ustore\ and in Git, we performed one commit operation with varying file
size.  We observe that \ustore\ is up to $200$x faster than Git, mainly due to Git's overhead when pushing commits to
the remote server. In particular,  before connecting to server, Git triggers object packing --- a time
consuming process --- that compressed multiple objects to save network bandwidth.

\subsection{Collaborative Data Science}


We evaluated Decibel against our implementation for the collaborative workflows (referred to as \ustore) on a
single-node setting, since Decibel does not support running on multiple nodes.  We first populated both
systems with a synthetic dataset similar to what was used in~\cite{decibel}. The dataset consists of $832,053$ tuples,
each has 10 integer fields with the first being the primary key. We used the science branching pattern, in
which new branches either start from the head of an active branch, or from the head commit of the master
branch. We set the page size in Decibel to $4$MB. We considered four queries as in~\cite{decibel}. The first
query (Q1) scans all the active tuples on master branch. The second query (Q2) scans all the active tuples on
the master branch but not on an another branch. The third query (Q3) scans all the tuples active on both 
the master and on another branch. The fourth query (Q4) scans all the tuples active in at least one branch. We
implement these queries in \ustore\ using the Version-First storage strategy. 

\begin{figure}
\centering
\includegraphics[width=0.3\textwidth]{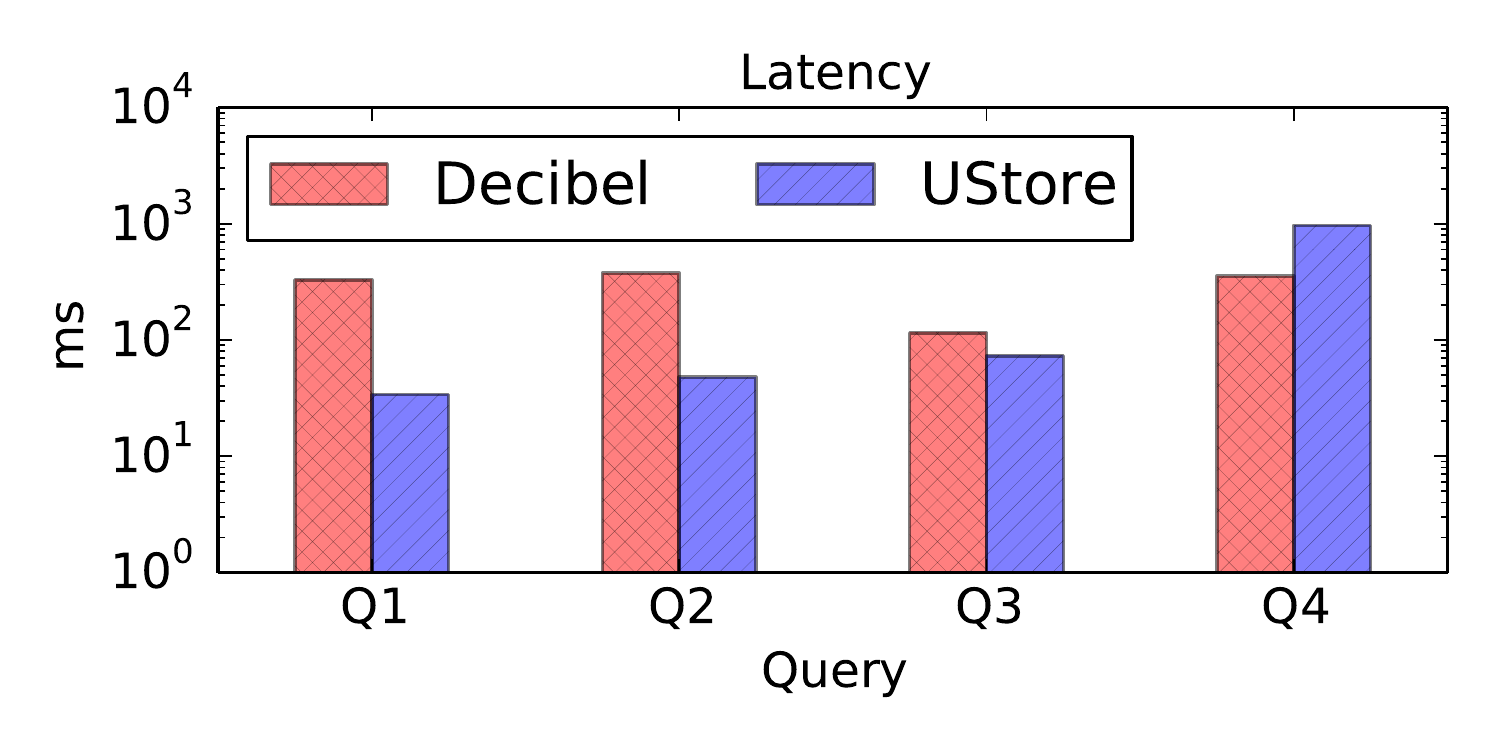}
\caption{Comparison Between Decibel and \ustore.}
\label{fig:DecibelVsUStore}
\end{figure}

Figure \ref{fig:DecibelVsUStore} compares the latency for the four queries. \ustore\ outperforms Decibel for
the first three queries. The performance gain is due to the fact that \ustore\ is memory-based, as opposed to
Decibel's disk-based storage backend. For example, \ustore\ takes only $33$ms to execute Q1, while Decibel
takes $328$ms. However, Decibel is better in Q4, because it implements optimizations based on branch topology
and with which it can avoid redundant scanning of common ancestors of different branches. \ustore\ has no such
optimizations and therefore takes longer to complete a full scan. We also compared both systems with the
Tuple-First storage strategy, and we observed that Decibel outperforms \ustore\ in most cases. This is due to another
optimization in Decibel where it can perform sequential access by scanning tuples in a single heap file for
each branch, while \ustore\ has to make many random accesses. We note that implementing these optimizations in \ustore\ is non-trivial, hence we plan to include them as part of future work.


\subsection{Transaction Management}
\begin{figure}
\centering
\includegraphics[width=0.27\textwidth]{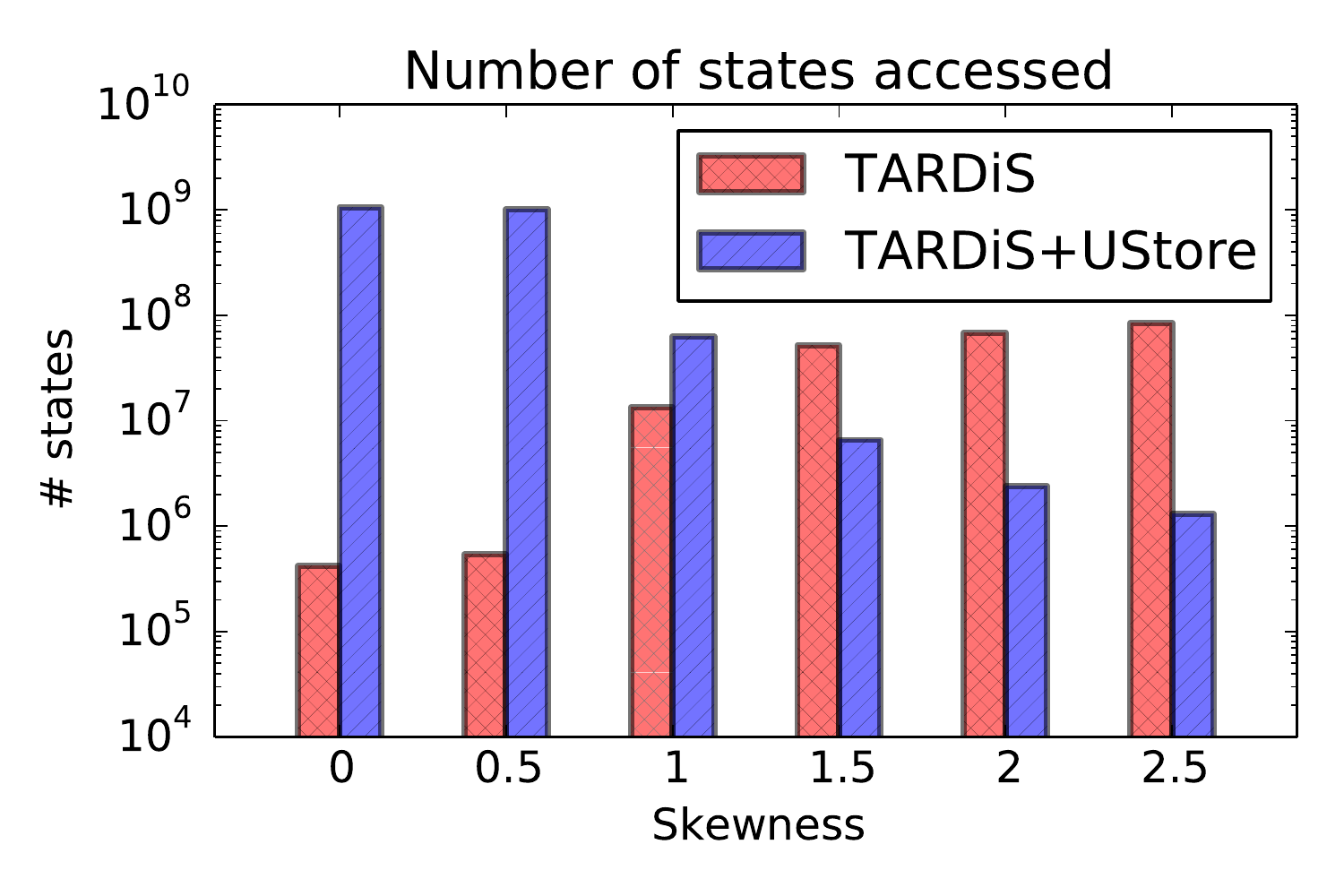}
\caption{Transaction management in TARDiS vs. \ustore.}
\label{fig:tardis}
\end{figure}
To evaluate TARDiS+\ustore\ against the original TARDiS, we used workloads consistent with what described
in~\cite{tardis}. In particular, a read-only transaction performs 6 reads, and a read-write transaction performs 3 reads and
3 writes. We considered two types of workloads: {\em read-heavy} which contains $75\%$ read-only transactions and 25\% read-write
transactions, and {\em mixed} which contains 25\% read-only transactions and 75\% read-write transactions.  We set the
number of concurrent clients to $1000$ and generated random data accesses following the Zipfian distribution.  We
report here the total number of state accesses by the read operations.  As the write operation is efficient in
both implementations, the performance gap is determined by the efficiency of read operations whose performance is
proportional to the number of state accesses.  

Figure~\ref{fig:tardis} shows the results for the mixed workload (those for the read-heavy workload are similar). It
can be seen that when the skewness of data accesses is low (e.g., Zipf=0 which is equivalent to uniform distribution),
TARDiS outperforms TARDiS+\ustore.  This is as expected, because TARDiS+\ustore\ has to scan longer branches to find the
versions.  But when the skewness is high (e.g., Zipf=1.5), TARDiS+\ustore\ shows clear advantages. This is because
updates of frequently accessed data can be found in recent states, therefore backtracking involves small numbers of hops
to locate the required versions.  When the skewness grows even higher, the performance gap widens accordingly.
In particular, when Zipf=$2.5$, the original TARDiS makes $80$x more state accesses than TARDiS+\ustore.  These results
demonstrate that our \ustore-based implementation achieves better performance when data accesses are skewed, which is a
common access pattern in real-world applications. 

\subsection{Blockchain}
\label{sec:eval_blockchain}

We compared \ustore-based blockchain implementation (referred to as \ustore) against Ethereum on queries
related to blockchain data. We generated a synthetic dataset with $1,000,000$ accounts in the genesis
block, and $1,000,000$ subsequent blocks containing 10 transactions per block on average. This dataset is
consistent with the public Ethereum data\footnote{https://etherscan.io/}.  We deployed both \ustore\ and
Ethereum's Geth client on a single node, and evaluated them over five queries. The first query (genesis block
load) is to parse and load the genesis block into the storage to initialize the system. The second query (general
block load) is to load all subsequent blocks into the storage. The third query (latest version scan) is to scan
the latest version of all the accounts.  The fourth query (block scan) is to scan the content of previous
blocks from a given block number. The fifth query (account scan) is to scan previous balances of a given
account at a given block. The last three queries represent the analytical workloads anticipated in the future
deployment of private blockchains~\cite{morgan16}. 

\begin{table}
\footnotesize
\centering
\caption{Ethereum performance vs. \ustore\ on block load and account scan}
\label{tab:gethvsustore}
\begin{tabular}{l|ll|l}
\hline
& \begin{tabular}[c]{@{}l@{}} {Genesis} \\ {Block}\\ {(sec)}\end{tabular} & \begin{tabular}[c]{@{}l@{}} {General} \\ {Block}\\ {(ms)}\end{tabular} & \begin{tabular}[c]{@{}l@{}}{Scan} \\ {Accounts}\\ {(sec)}\end{tabular} \\ \hline
{Ethereum} & 622.45 & \textbf{19} & 1148.096 \\
\multicolumn{1}{r|}{{\ustore}} & \textbf{175.516} & 35 & \textbf{167.546} \\ \hline
\end{tabular}
\end{table}


\begin{figure}
\centering
\includegraphics[width=0.36\textwidth]{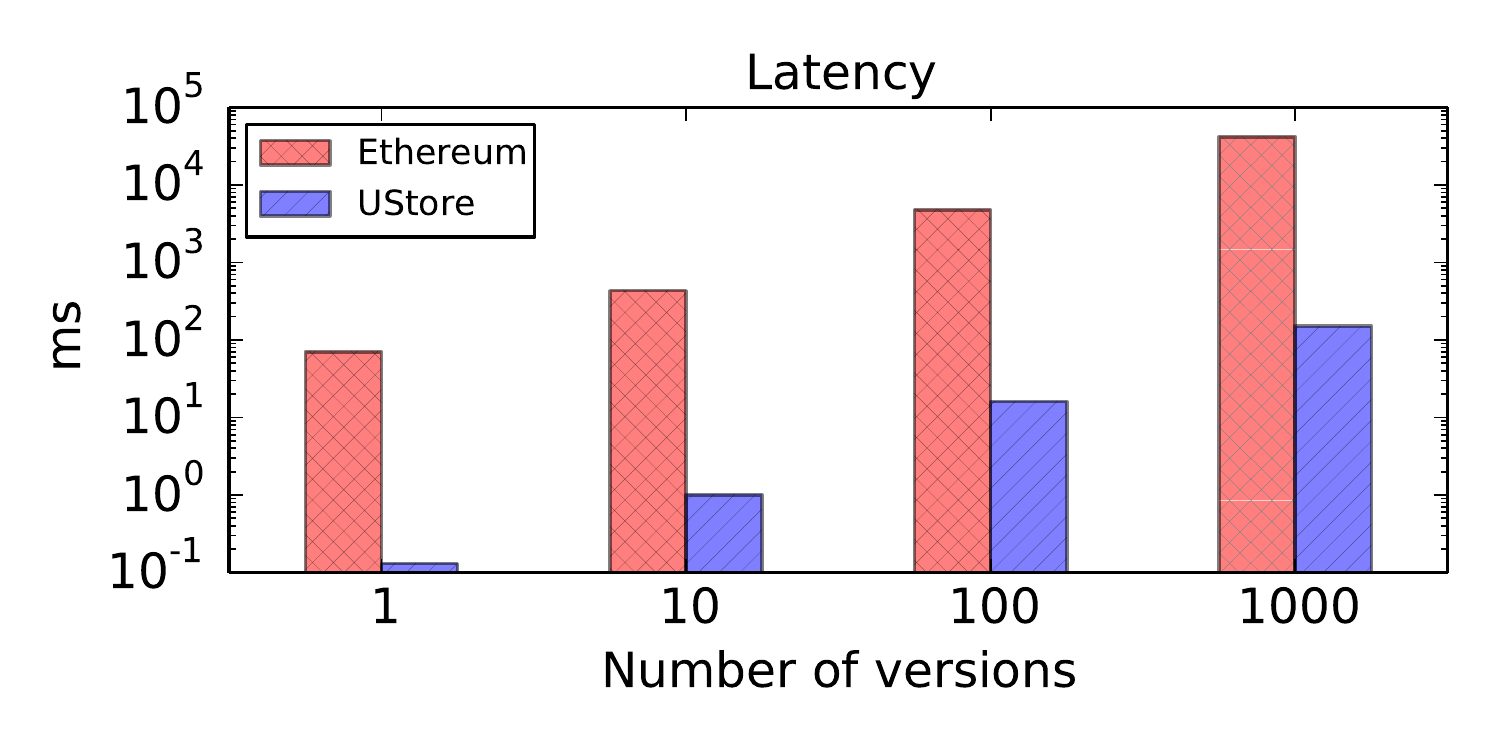}
\caption{Ethereum performance vs. \ustore\ on account level version scan}
\label{fig:account_version_scan}
\end{figure}

Table \ref{tab:gethvsustore} summarizes the latency of the first three operations in Ethereum and in \ustore.
Figure \ref{fig:account_version_scan} shows the latency for the account scan query, in which \ustore\ is up to 
$400$x better (the results for block scan query are similar and included in the Appendix). It can be seen that
\ustore\ outperforms Ethereum on all operations except the general block loading operation. The performance gain
is attributed to three factors. One factor is that \ustore\ serves data from memory, and it leverages RDMAs.
Another factor is that it uses an AVL variant\footnote{https://github.com/tendermint/go-merkle} to index
the accounts, which has better performance for insertion than Ethereum's Patricia-Merkle tree. But more
importantly, \ustore's data structures are a better match for version oriented queries than Ethereum's, and
thus the application can directly benefit from the efficient version operations provided by the storage. For
instance, \ustore\ is better for the last two queries because of the efficient version scan operation.
Furthermore, for the account scan query, Ethereum maintains no explicit pointers to the previous version of an
account, thus it has to fetch and parse the content of a previous block before retrieving the previous
account version. This process requires reading redundant blocks and therefore incurs more overhead.  In
contrast, \ustore\ can leverage the {\tt GetKPreviousVersion/Value} API to retrieve the previous versions
directly.  Ethereum is better for general block load query because it imports blocks in batches without
verifying the content, whereas \ustore\ imports one by one.


\section{Discussion and Future Work}
\label{sec:discussion}

The current evaluation of \ustore\ focuses only on the system's core APIs. View-layer components, especially the access
control and notification service, are still to be evaluated in isolation and as parts of the overall performance.
Furthermore, the applications in \ustore\ (except for the transaction management application) were compared against
full-fledged systems which support many others operations beside what were implemented in \ustore. In other words, we
ported only a small number of features from these systems into \ustore, ignoring many others that may contribute to the
overall performance. Adding more features to these applications are part of future work. 

Much of the future work, however, is to enhance the view layer. The current access control mechanism is restricted to
read policies, and it assumes trusted servers. Supporting fine-grained write access is non-trivial, since the semantics of
write access to a version needs to be formally defined, and the enforcement must be space efficient. Even for read
policy, it remains a challenge to compactly represent a policy concerning multiple versions, since version numbers are
random. The next step is to relax the trust assumption, for which we plan to exploit trusted hardware (Intel SGX) to run
the enforcement protocol in the trusted environment. 

A view-layer module that implements more advanced data models from the core abstraction can greatly reduce development
effort. For example, our implementation of Git and collaborative applications require several levels of \uobject s in
which one level stores pointers to another. Such grouping of low-level \uobject s into higher-level abstractions is
useful to track provenance and changes application meta-data~\cite{ground}. Most applications we ported to \ustore\
require more complex access to the data than the current {\tt Put} and {\tt Get} APIs. Thus, we plan to implement a
query and analytics engine as a view-layer module to support rich operations over the immutable data. Other interesting
modules that enrich the view layer include utility modules that support authentication and integration with existing
systems (like a big-data pipeline or a machine learning system). 

The current physical layer is designed for data-center environments. Extending it to non-cluster (decentralized)
settings requires addressing two challenges: disk-based storage management (garbage collection), and data management in
the presence of churn. At the other end of the stack, we note a recent rise of big meta-data systems. As more systems
are being built around interactions with human experts to extract contextual intelligence~\cite{icrowd}, capturing the
context of such interactions is crucial for improving the systems. Furthermore, as machine learning models are becoming
important for decision making and system tuning~\cite{selfdrivingdbms}, understanding their provenance is an important
step towards better interpretation of the models.  Ground~\cite{ground} and Goods~\cite{goods} are two meta-data systems
focusing on extracting and managing data context, especially data provenance.  One central component in Ground is the
immutable, multi-version storage system, which the authors demonstrated that existing solutions fall short.  \ustore\
can be an integral part in such systems, and consequently serve the emerging classes of applications based on meta-data.  

\section{Conclusion}
\label{sec:conclusion}
In this paper, we identified three properties commonly found in many of today's distributed applications,
namely data immutability, sharing and security. We then designed a flexible, efficient and scalable
distributed storage system, called \ustore, which is enriched with these properties. We demonstrated how the new
storage adds values to existing systems and facilitates faster development of new applications by implementing
and benchmarking four different applications on \ustore.  
\newline

\section*{Acknowledgements}
This work was supported by the National Research Foundation, Prime Minister's Office, Singapore under Grant number
NRF-CRP8-2011-08.

\bibliographystyle{abbrv}

\bibliography{paper} 
\appendix

\begin{figure}[h]
\centering
\includegraphics[width=0.5\textwidth]{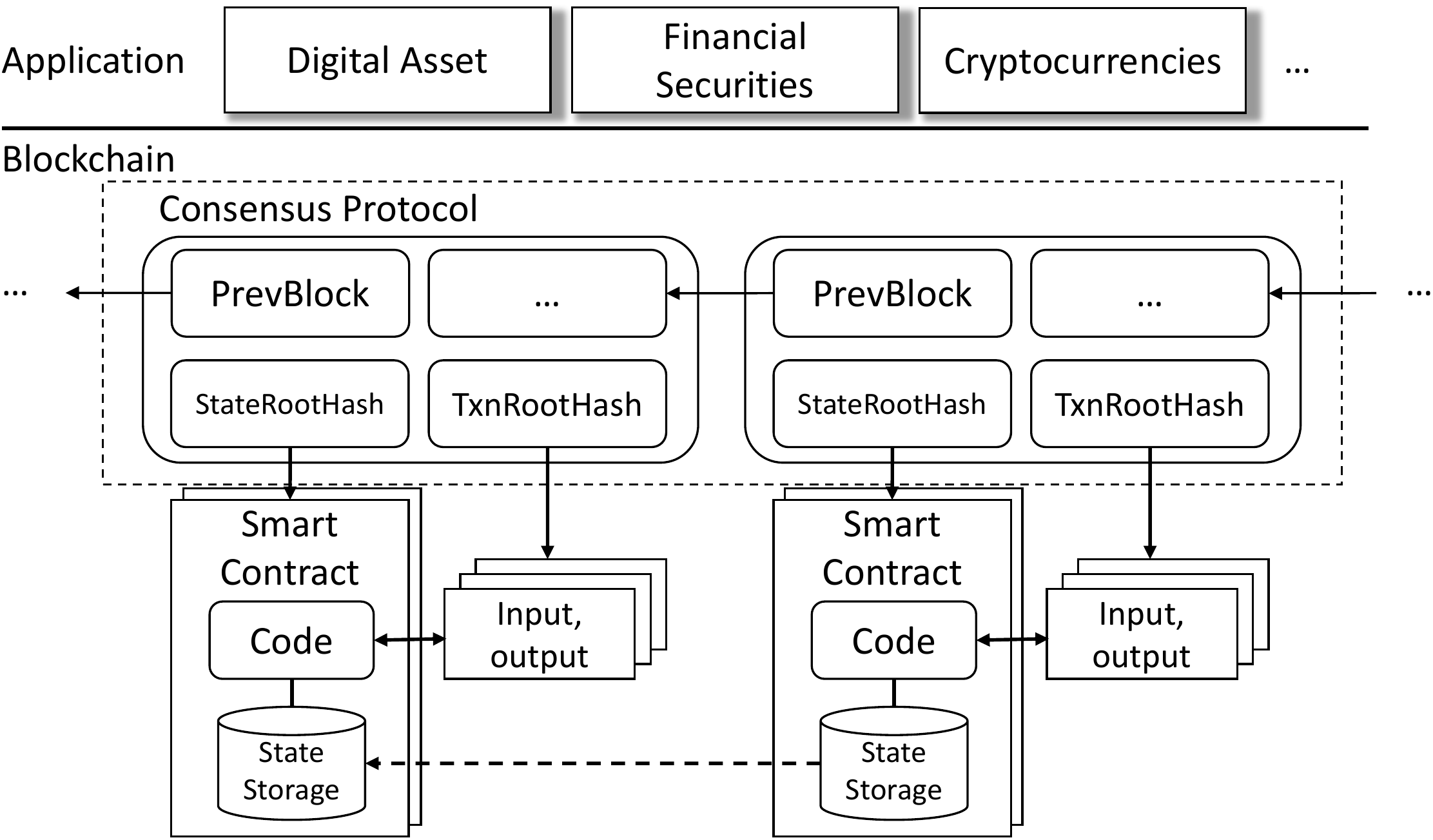}
\caption{Blockchain software stack}
\label{fig:blockchain_architecture}
\end{figure}

\section{UStore Applications}
\subsection{Advanced Git Application}
The simple Git implementation described in Section~\ref{sec:newapp} only tracks version history at commit level as in
original Git protocol.  To extend this, we target to track histories for all data types, especially blob (file) and tree
(folder).  The immediate benefit is that  users can view change list of a file or folder more quickly.

To achieve this, we need to separate files and folders as different \uobject s, using its their name as keys and
serialized content as value.  When we detect that the content has changed during a commit, (e.g., same file name but
different content as in previous checkout) we invoke {\tt Put}$(\textit{name}, \textit{version}, \textit{content})$.

The merge operation for two versions is similar.  After resolving conflicts and generating the merged content, we invoke
{\tt Merge}$(\textit{name}, \textit{version1}, \textit{version2}, \textit{content})$.  This returns a version id which can be
used to fetch that object.

\subsection{Blockchain}

We only described the data structure of blockchain in Section~\ref{sec:app_blockchain}, but a typical blockchain system consists of many more complex components besides the data storage. Here we elaborate a typical architecture of a blockchain system as shown in Figure~\ref{fig:blockchain_architecture}.

Many applications such as cryptocurrencies, digital asset management and finical security settlement can built upon blockchains. Typically, some of the nodes in a blockchain network may exhibit Byzantine behavior, but the majority is honest, thus, the nodes have to reach agreement on the unique evolution history of the global system state through a consensus protocol (e.g., PBFT in Hyperledger\cite{hyperledger}, PoW in Bitcoin\cite{bitcoin} and Ethereum\cite{ethereum}). To be general and extensible, recent blockchain platforms, like Ethereum and Hyperledger, support smart contract functionality that allows users to define their own transaction logic and develop their softwares called decentralized application (DApp), e.g. decentralized online money exchange. The smart contracts are usually written in Turing-complete languages and executed in an isolated execution environment provided by the blockchain, such as Ethereum Virtual Machine (EVM) of Ethereum. Because the nodes in blockchain network cannot be fully trusted, blockchain platforms usually construct validity proofs based on Merkle hash tree and its variants that are also used as the indexes for system internal states. 

Although the storage engine is only a part of the complex blockchain software stack, there is a need for efficiency and scalability which is critical for smart contract execution and historical data audit.
As we shown in our experiment, \ustore\ can serve as the data store of choice for high performance blockchains.

\section{Experimental Results}
\subsection{Microbenchmark}
\begin{figure}
\centering
\includegraphics[width=0.35\textwidth]{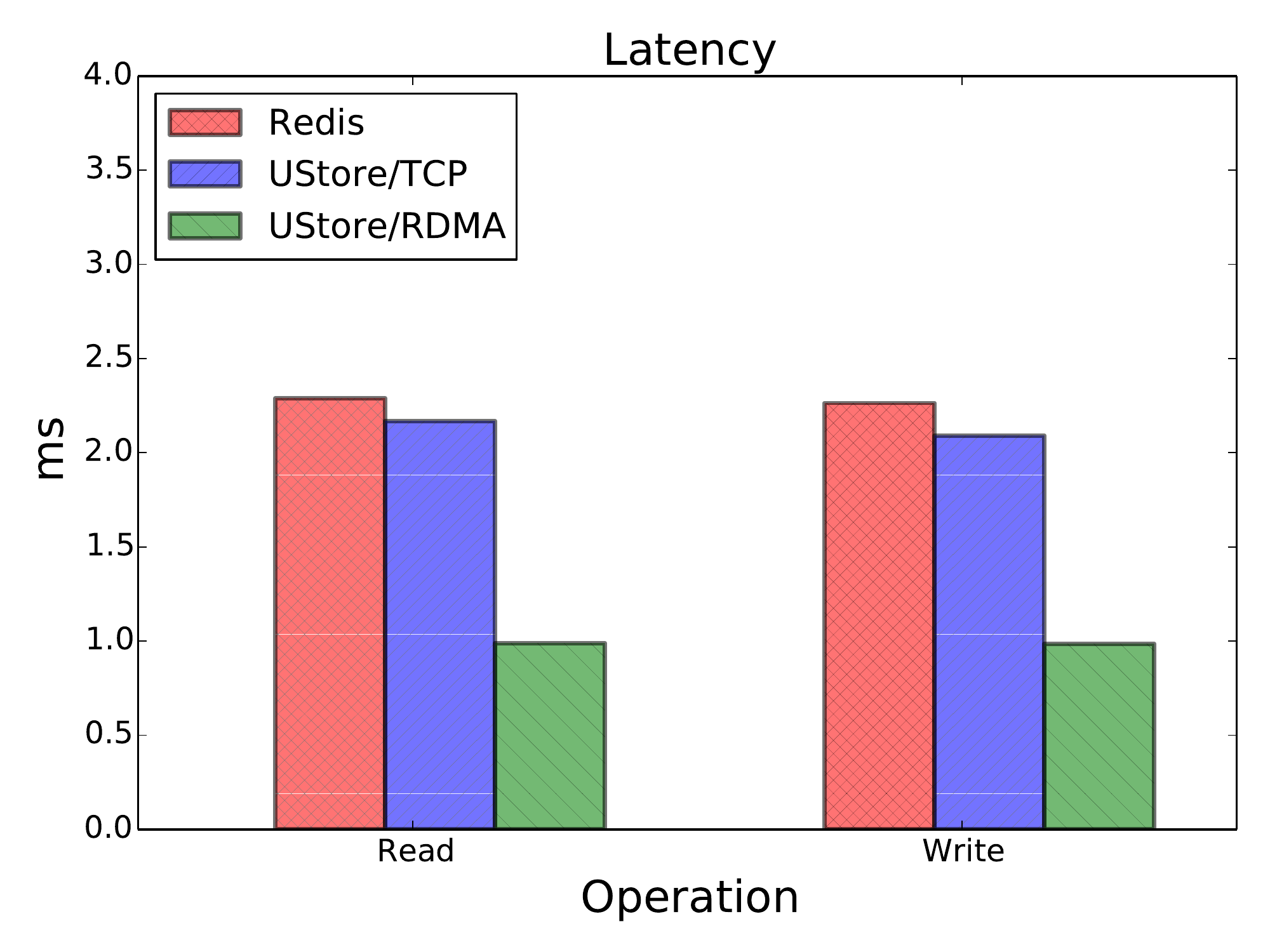}
\caption{Latency vs. Redis.}
\label{fig:lt}
\end{figure}

\begin{figure}
  \centering
  \vspace{-8pt}
  \begin{subfigure}[b]{0.4\textwidth}
  \centering
     	\includegraphics[width=\linewidth]{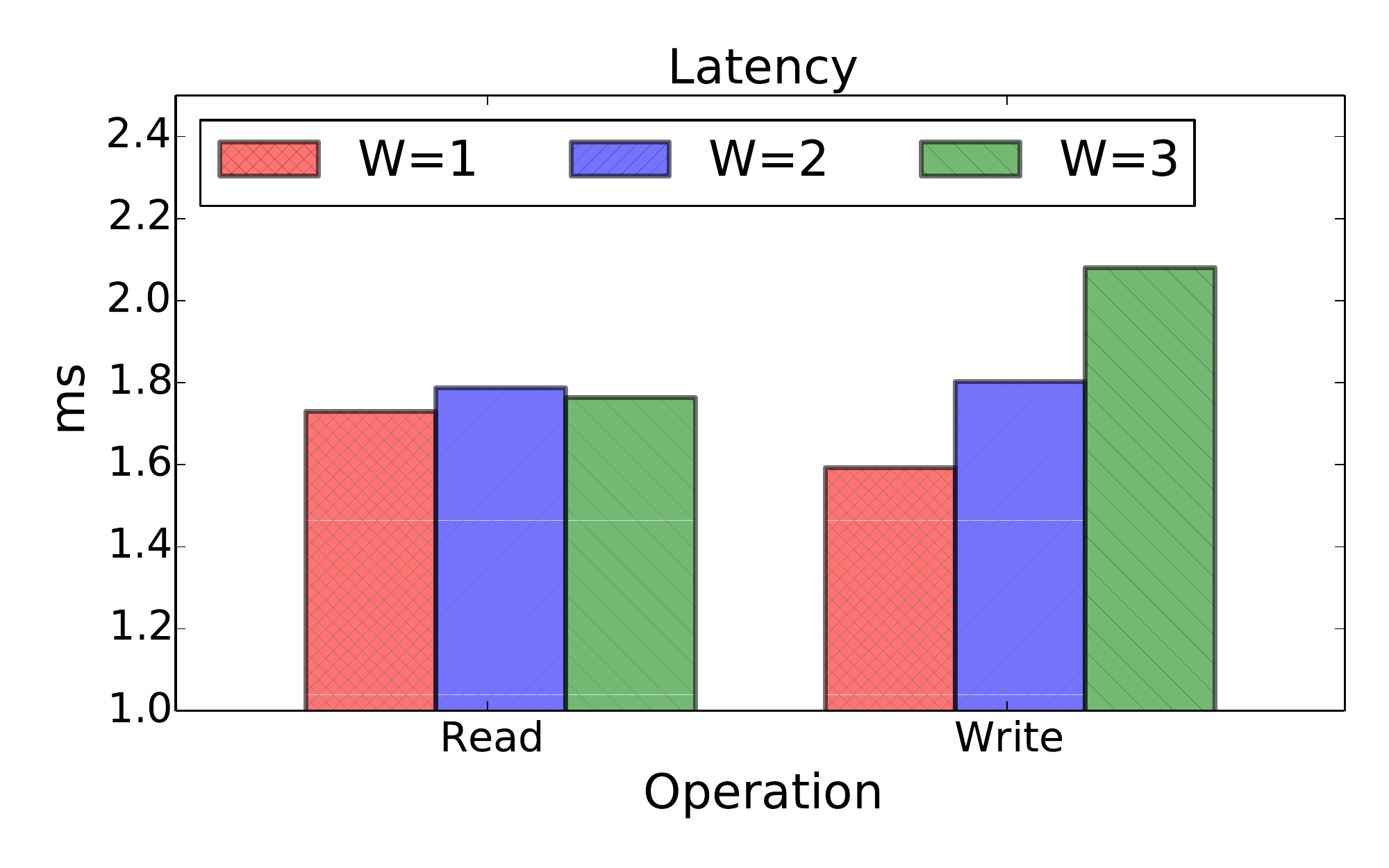}
     	\caption{Latency}
     	\label{fig:w-lt}
  \end{subfigure}
  \\
  \vspace{10pt}
  \begin{subfigure}[b]{0.4\textwidth}
  \centering
     	\includegraphics[width=\linewidth]{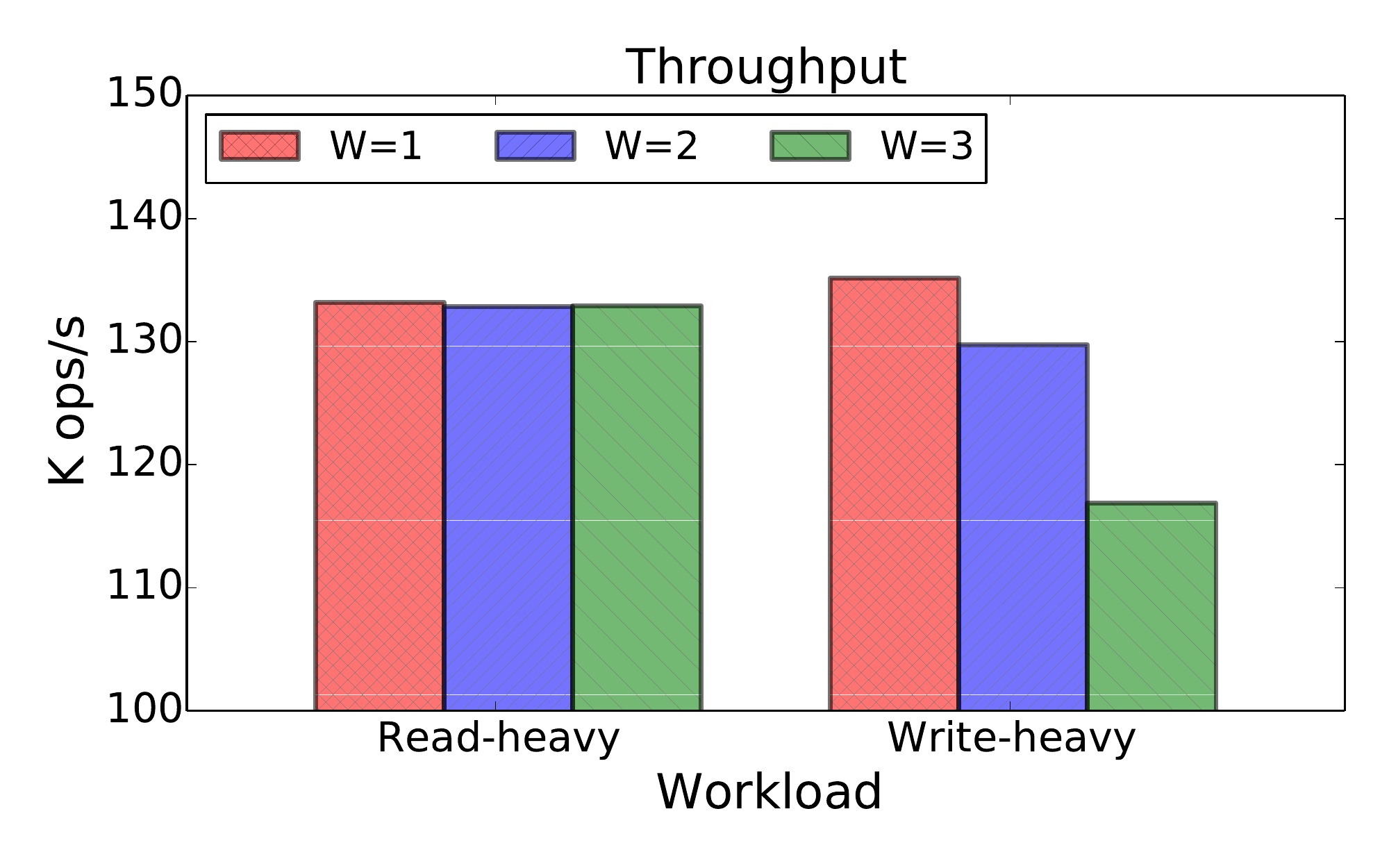}
     	\caption{Throughput}
     	\label{fig:w-tp}
  \end{subfigure}
  \caption{Performance trade-offs against the number of write replicas $W$.}
  \label{fig:w_lt}
  \vspace{-4pt}
  \end{figure}

Figure~\ref{fig:lt} shows the latency for read and write operations with 4 clients. It can be seen that \ustore\ can
achieve $3$x lower latency thanks to RDMAs. Even without RDMAs, \ustore\ is slightly better than Redis, because our
server is multi-threaded. Another source of overhead can be due to Redis' support for complex data types such as
lists and sorted sets, whereas \ustore\ supports simple, raw binary format.   

Figure~\ref{fig:w_lt} illustrates \ustore's performance with varying number of write replicas $W$. For this experiment, we
run 4 \ustore\ nodes and set $N=3$. Recall that $W$ determines how many responses from the replicas are needed before a
write operation is considered successful. As a result, higher $W$ leads to higher latency for writes, as we can see for
write-heavy workloads in Figure~\ref{fig:w_lt}. $W$ only affects read operations in the extreme cases where a read
request is sent immediately after the write request to a replica which has not finished writing the data, which is rare
in our cases. Thus, we observed no impact of $W$ on read-heavy workloads. 

    \begin{figure}
      \centering
      \vspace{-8pt}
      \begin{subfigure}[b]{0.34\textwidth}
      \centering
      	\includegraphics[width=\linewidth]{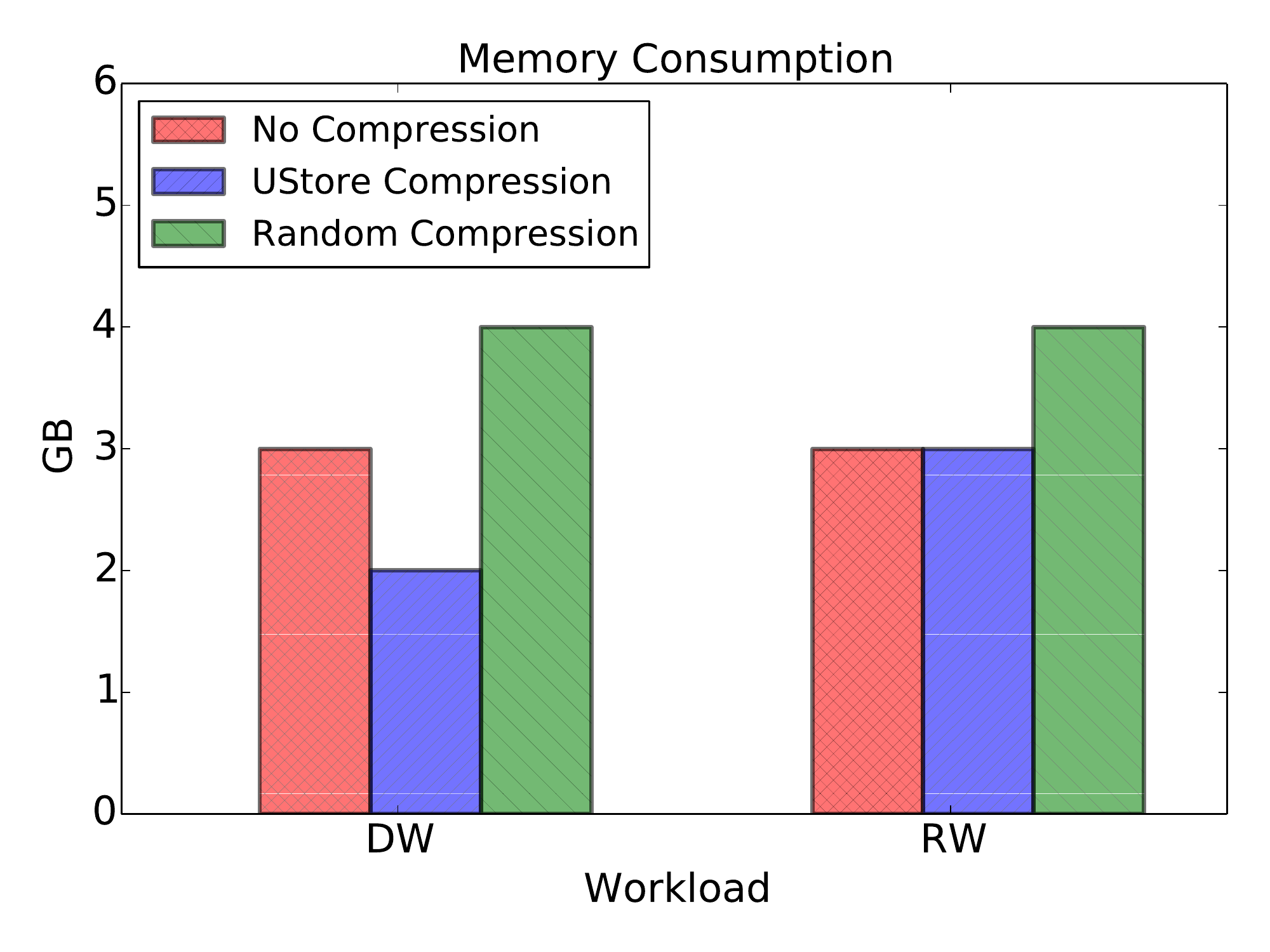}
      	\caption{Memory}
      	\label{fig:comp-mem}
      \end{subfigure}
      \\
      \vspace{10pt}
      \begin{subfigure}[b]{0.37\textwidth}
      \centering
      	\includegraphics[width=\linewidth]{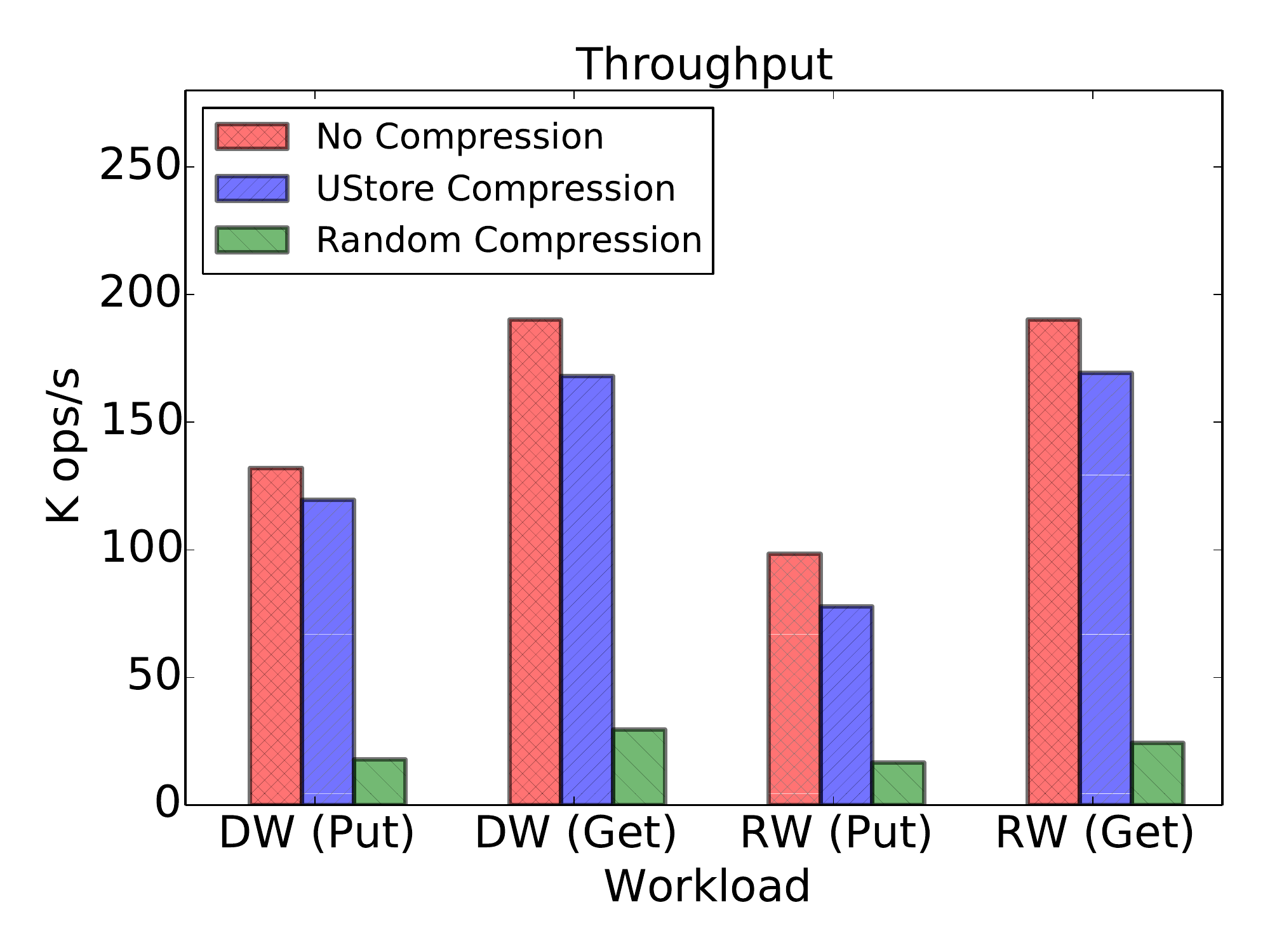}
      	\caption{Throughput}
      	\label{fig:comp-thr}
      \end{subfigure}
      \caption{Compression Benchmark}
      \label{fig:comp}
      \vspace{-4pt}
      \end{figure}
{\bf Compression strategies.}
We benchmarked the effect of our compression strategy,
by comparing it with the \textit{no compression} strategy and a \textit{random compression} strategy,
where we randomly select an ancestor
based on which we do the compression,
in terms of both the performance and memory consumption.
We generated two workloads with 1.6 M records of 1 KB length and long version traces,
i.e., a delta workload (DW)
where the current version is generated based on its previous version
with some delta changes,
and a randomly generated workload (RW) where every version is generated randomly.
We used 2 servers and 16 clients.
The results are shown in Figure~\ref{fig:comp}.
We can see that under the delta workload (DW), 
our compression strategy saves a lot of memory space,
but only performs partially worse than the \textit{no compression} strategy,
while under the random workload (RW),
their memory consumption is similar,
because there is no much opportunity for effective compression under random workload (RW). 
The \textit{random compression} strategy
performs worst,
in terms of both memory consumption and performance under both workloads,
which is mainly due to the high probability of
network communication during the compression and de-compression,
and it does not take advantage of the fact that
only neighboring versions are likely to have similar content
(i.e., there are overlooked opportunities for effective compression).


\subsection{Git}
\begin{figure}
\centering
\includegraphics[width=0.37\textwidth]{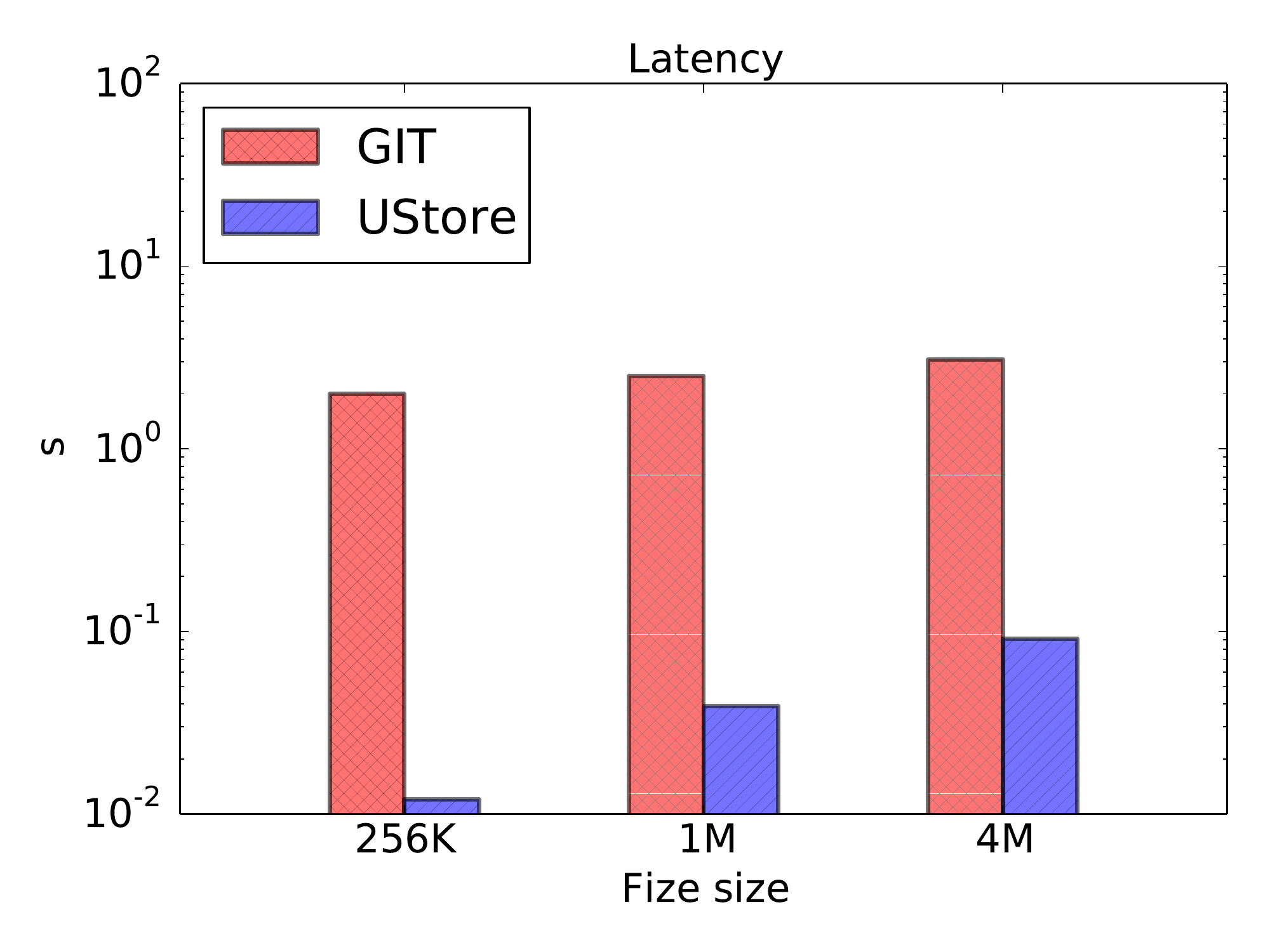}
\caption{Commit operations in Git vs. {\ustore}.}
\label{fig:git_commit}
\end{figure}

\begin{figure}
\centering
\includegraphics[width=0.37\textwidth]{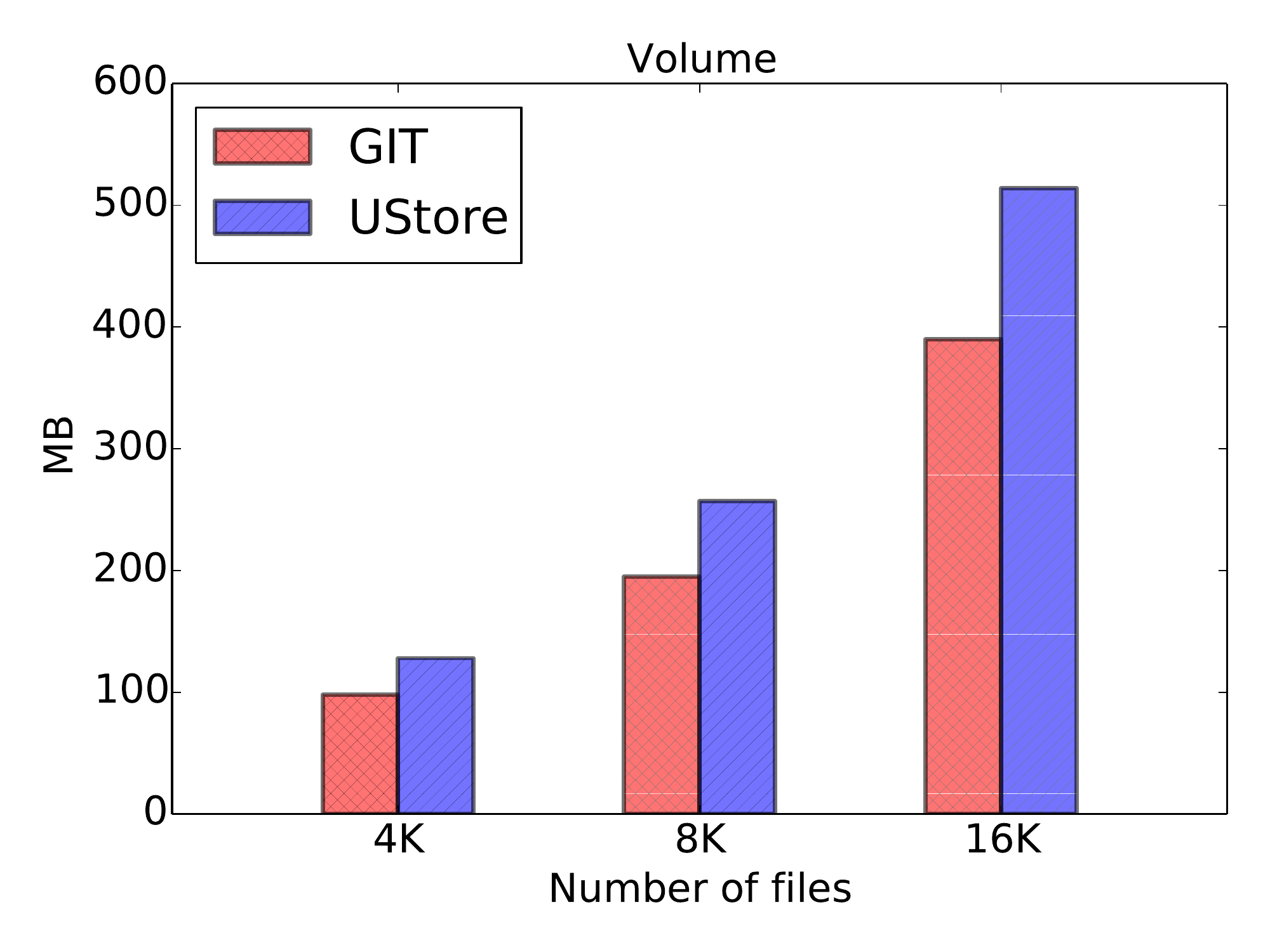}
\caption{Storage consumption in Git vs. \ustore\ for normal (unencrypted) workload}
\label{fig:git_space_plain}
\end{figure}

\begin{figure}
\centering
\includegraphics[width=0.37\textwidth]{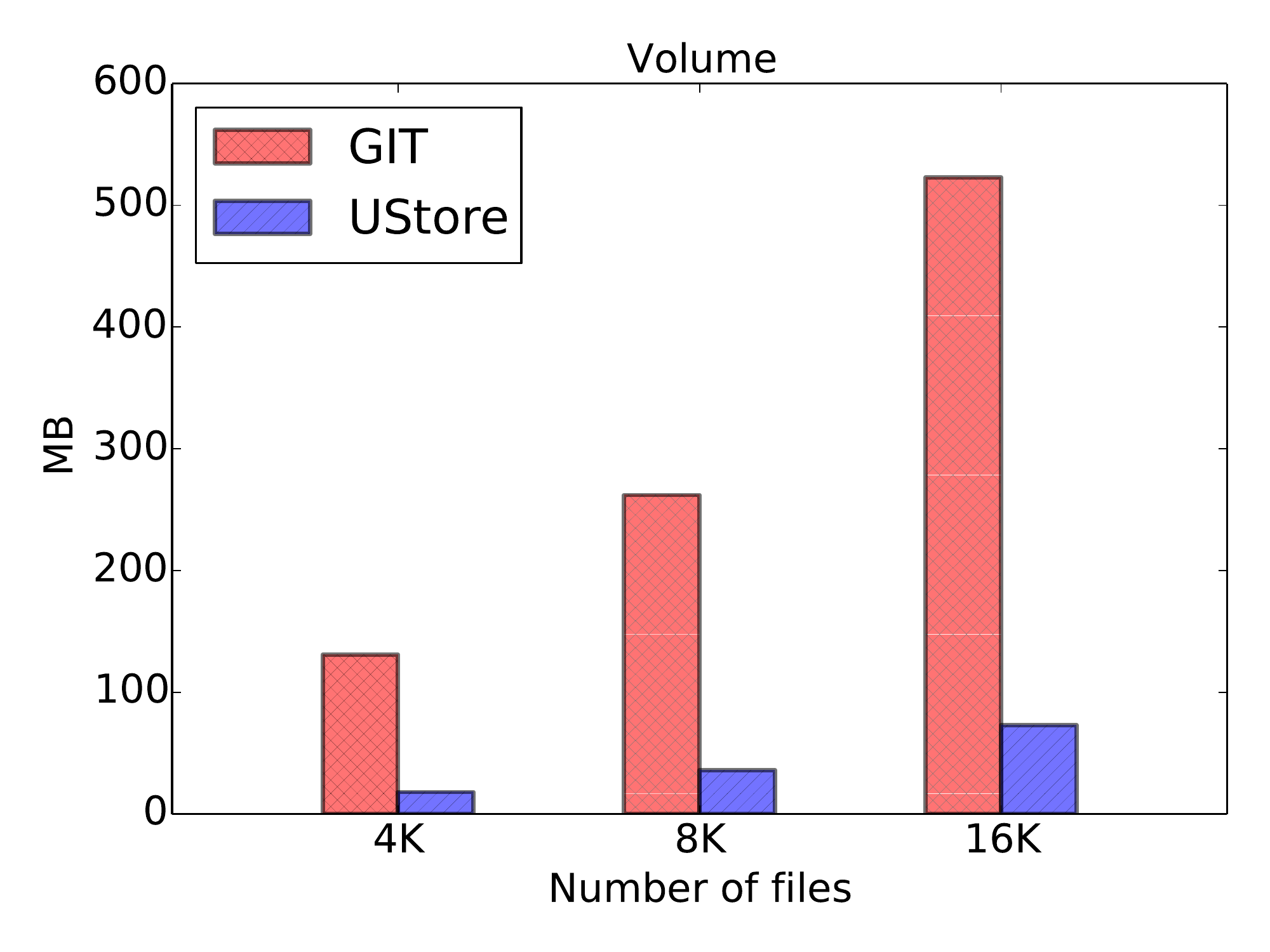}
\caption{Storage consumption in Git vs. \ustore\ for encrypted workload}
\label{fig:git_space_encrypt}
\end{figure}

Figure~\ref{fig:git_commit} illustrates the cost for committing a file into Git versus into \ustore.
It can be seen that the commit operation in \ustore\ is up to 200x faster than in Git. This performance gap is
attributed to several factors.  First, \ustore\ stores data in memory which avoids external I/Os.  Second, Git
incurs much overhead when pushing commits to the remote server.  Before connecting to server, Git automatically
triggers object packing to combine multiple objects as compressed packages.  This strategy reduces network
communication, but it is time consuming.

Figure~\ref{fig:git_space_plain} and Figure~\ref{fig:git_space_encrypt} show the storage space consumption for
maintaining a Git repository at the server side.  In order to reduce space consumption, Git uses {\em zlib} to
compress all objects, reducing both storage and network cost.  We conducted a test with plain text workload.
As shown in Figure~\ref{fig:git_space_plain}, Git occupies less space than \ustore, since we do not apply any
compression in \ustore.  However, \ustore\ exposes the interface for users to define own compression
strategies, e.g. {\em zlib}.  More importantly, compression in \ustore\ can be application-specific.  Users
can leverage unique characteristics of their data to perform much better.  For example, we generated an AES encrypted
workload, which is ineffective to compress directly. We then injected a customized compression function
into \ustore\ which processes the data in decrypt-compress-encrypt manner.  Figure~\ref{fig:git_space_encrypt}
validates the effectiveness of this application-specific compression.  We can see that if users know their
data well, it is easy for \ustore\ to outperform the default compression in Git.

\subsection{Blockchain Application}
\begin{figure}
\centering
\includegraphics[width=0.4\textwidth]{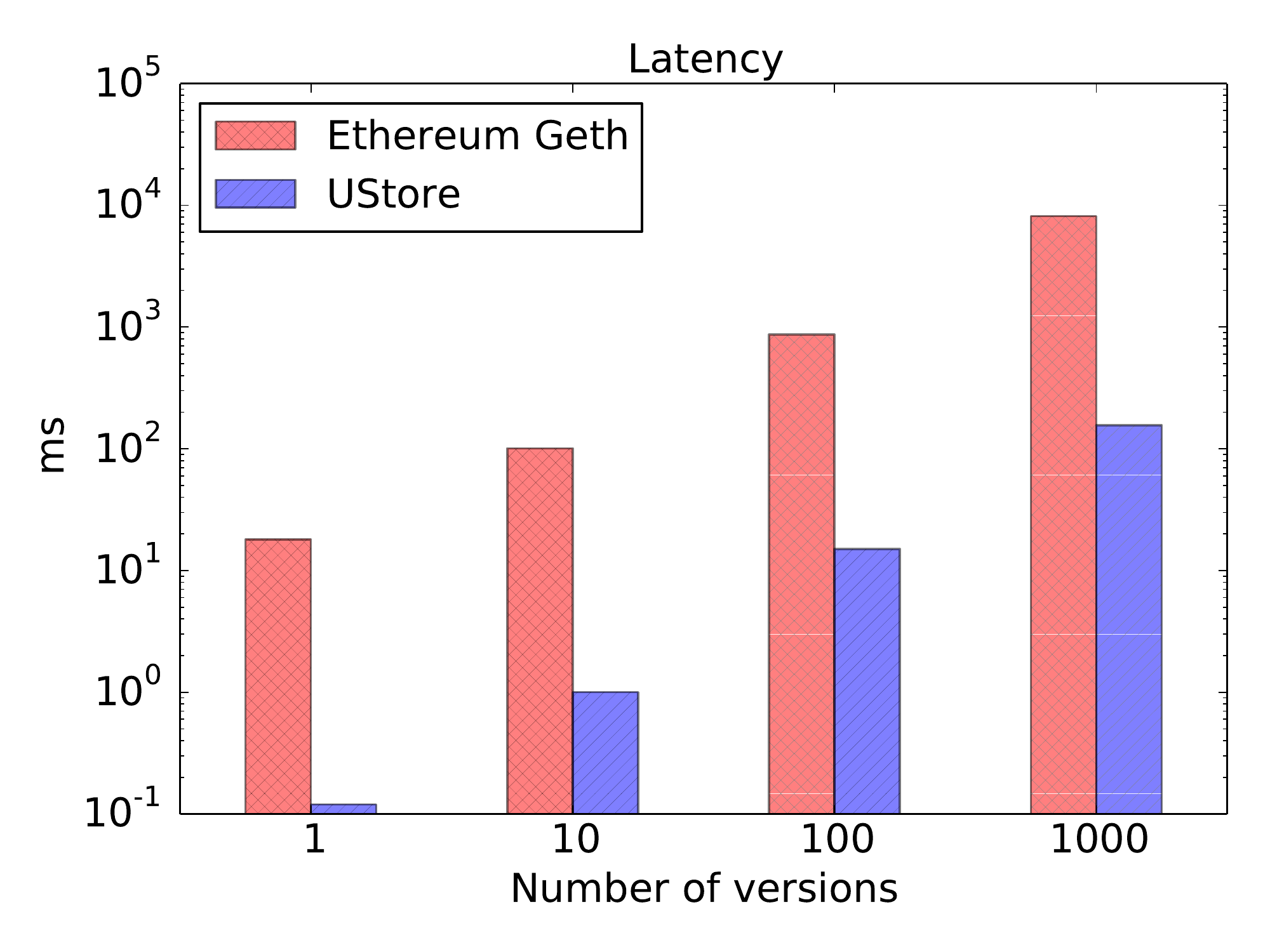}
\caption{Ethereum performance vs. \ustore\ on block level version scan}
\label{fig:block_version_scan}
\end{figure}
Figure~\ref{fig:block_version_scan} illustrates comparison between Geth and \ustore\ on block level version scan operations
described in Section \ref{sec:eval_blockchain}. It shares a similar performance pattern with account level version scan operation shown in Figure~
\ref{fig:account_version_scan}, but incurs less overhead in Geth than account level version scan. This is because in Ethereum's implementation, when scanning account versions, it has to fetch the block information corresponding to that account version as well.

\end{document}